\documentclass[aps,twocolumn,prd,floatfix,showpacs,10pt]{revtex4-1}
\usepackage{amsmath} \usepackage{amsfonts} \usepackage{amssymb}
\usepackage{amsthm} \usepackage{mathrsfs} \usepackage{multirow}
\usepackage{color} \usepackage[caption=false]{subfig}
\usepackage{array} \usepackage{subfig} \usepackage{tikz}

\usepackage[breaklinks=true, colorlinks=true, urlcolor=blue,
citecolor=blue, link color=black]{hyperref}

\def\dsp{\displaystyle}
\def \Re{\text{Re}} \def \Im{\text{Im}} 

\def\nn {\nonumber} \def\sss{\scriptscriptstyle}

\def\gev{\ensuremath{\mathrm{Ge\kern -0.1em V}}}
\def\tev{\ensuremath{\mathrm{Te\kern -0.1em V}}}
\def\gammaf{\ensuremath{\mathrm{\Gamma\kern -0.2em _{f}}~}}

\newcommand{\modulus}[1]{\left| #1 \right|}

\allowdisplaybreaks

\begin{document}
\urlstyle{rm}

\title{Inferring the nature of the boson at 125-126 {Ge\kern -0.1em V}}

\author{Arjun Menon} \email{aamenon@uoregon.edu}
\affiliation{Institute of Theoretical Science, University of Oregon}
\author{Tanmoy Modak} \email{tanmoyy@imsc.res.in}
\author{Dibyakrupa Sahoo} \email{sdibyakrupa@imsc.res.in}
\author{Rahul Sinha} \email{sinha@imsc.res.in} \affiliation{The
  Institute of Mathematical Sciences, Taramani, Chennai 600113, India}
\author{Hai-Yang Cheng} \email{phcheng@phys.sinica.edu.tw}
\affiliation{Institute of Physics, Academia Sinica, Taipei, Taiwan
  11529, Republic of China}

\date{\today}

\begin{abstract}
  The presence of a bosonic resonance near 125~\gev~has been firmly
  established at the Large Hadron Collider. Understanding the exact
  nature of this boson is a priority. The task now is to verify
  whether the boson is indeed the scalar Higgs as proposed in the
  Standard Model of particle physics, or something more esoteric as
  proposed in the plethora of extensions to the Standard Model. This
  requires a verification that the boson is a $J^{PC}=0^{++}$ state
  with couplings precisely as predicted by the Standard Model. Since a
  non Standard Model boson can in some cases mimic the Standard Model
  Higgs in its couplings to gauge bosons, it is essential to rule out
  any anomalous behavior in its gauge couplings.  We present a step by
  step methodology to determine the properties of this resonance
  without making any assumptions about its couplings.  We present the
  analysis in terms of uni-angular distributions which lead to angular
  asymmetries that allow for the extraction of the couplings of the
  125-126~\gev~resonance to Z bosons.  We show analytically and
  numerically, that these asymmetries can unambiguously confirm
  whether the new boson is indeed the Standard Model Higgs boson.
\end{abstract}


\maketitle

\section{Introduction}\label{sec:intro}

A new bosonic resonance with a mass of about 125~\gev~has recently
been observed at the Large Hadron Collider by both ATLAS Collaboration
\cite{:2012gk, ATLAS:science} and CMS Collaboration \cite{:2012gu,
  CMS:science, :2012br}. The mass of the resonance is suggestive that
this resonance is the Higgs boson that should exist in the Standard
Model of particle physics as a spin zero parity-even
resonance. Significant effort is now directed at determining the
properties and couplings of this new resonance to confirm that it is
indeed the Higgs boson of the Standard Model.  In this work we specify
this new boson by the symbol $H$ and we call it the Higgs, even though
it has not been proved to be the Higgs of the Standard Model. This
resonance is observed primarily in three decay channels $H\to
\gamma\gamma$, $H\to ZZ$ and $H\to WW$, where one (or both) of the
$Z$'s and $W$'s are off-shell. It is well known that the spin and
parity of the resonance and its couplings can be determined by
studying the momentum and angular distributions of the decay
products. Indeed there is little doubt that a detailed numerical fit
to the invariant masses of decay products and their angular
distributions will reveal the true nature of this resonance. However,
a detailed study of the angular distributions requires large
statistics and may not be feasible currently. Several studies existed
in the literature before the discovery of this new resonance
\cite{Nelson:1984bb,Dell'Aquila:1985ve,Nelson:1986ki,Kramer:1993jn,
  Barger:1993wt, Gunion:1996xu,
  Miller:2001bi,Bower:2002zx,Choi:2002dq, Choi:2002jk, Godbole:2002qu,
  Buszello:2002uu,Desch:2003mw,Worek:2003zp,Kaidalov:2003fw,
  Godbole:2004xe,Buszello:2006hf,Przysiezniak:2006fe,Bluj:2006,
  BhupalDev:2007is,Godbole:2007cn,Godbole:2007uz,Gao:2010qx,
  Englert:2010ud,Eboli:2011bq,DeSanctis:2011yc,Berge:2011ij,
  Kumar:2011yta,Ellis:2012wg,Englert:2012ct, Bredenstein:2006rh} and
yet several papers have appeared recently on strategies to determine
the spin and parity of the resonance \cite{Ellis:2012jv, Ellis:2012xd,
  Giardino:2012dp, Choi:2012yg, Boughezal:2012tz, Banerjee:2012ez,
  Avery:2012um, Coleppa:2012eh,Geng:2012hy, Ellis:2012mj,Frank:2012wh,
  Djouadi:2013-21-Jan,Englert:2012xt,Stolarski:2012ps}.  Yet, there is
no clear conclusion on the step by step methodology to determine these
properties and convincingly establish that the new resonance is indeed
the Standard Model Higgs boson. The recent result~\cite{:2012br} from
CMS Collaboration on the determination of spin and parity of the new
boson is not conclusive.

In this paper we are exclusively concerned with Higgs decaying to four
charged leptons, which proceeds via a pair of $Z$ bosons: $H \to ZZ
\to (\ell_1^- \ell_1^+) (\ell_2^- \ell_2^+),$ where $\ell_1$, $\ell_2$
are leptons $e$ or $\mu$. Since the Higgs is not heavy enough to
produce two real $Z$ bosons, we can have one real and another
off-shell $Z$, or both the $Z$'s can be off-shell. While we deal with
the former case in detail our analysis applies equally well to the
later case. We find that only in a very special case dealing with
$J^P=2^+$ boson it is more likely that both the $Z$ bosons are
off-shell. We emphasize that the final state $(e^+e^-)(\mu^+\mu^-)$ is
not equivalent to $(e^+e^-)(e^+e^-)$ or $(\mu^+\mu^-)(\mu^+\mu^-)$ as
sometimes mentioned in the literature, since the latter final states
have to be anti-symmetrized with respect to each of the two sets of
identical fermions in the final state. The anti-symmetrization of the
amplitudes is not done in our analysis and hence our analysis applies
only to $(e^+e^-)(\mu^+\mu^-)$.  We examine the angular distributions
and present a strategy to determine the spin and parity of $H$, as
well as its couplings to the $Z$-bosons with the least possible
measurements. Since the decay mode $H \to \gamma\gamma$ has been
observed, $H$ is necessarily a boson and the Landau-Yang
theorem~\cite{Landau:1948kw, Yang:1950rg} excludes that it has spin
$J=1$. Further, assuming charge conjugation invariance, the
observation of $H\to \gamma\gamma$ also implies~\cite{Barger:1993wt}
that $H$ is a charge conjugation $C=+$ state. In making this
assignment of charge conjugation it is assumed that $H$ is an
eigenstate of charge conjugation.  With the charge conjugation of $H$
thus established we will only deal with the parity of $H$
henceforth. We consider only Spin-0 and Spin-2 possibilities for the
$H$ boson.  Higher spin possibilities need not be considered for a
comparative study as the number of independent helicity amplitudes
does not increase any more \cite{Kramer:1973nv,Choi:2002jk}. The
process under consideration requires that Bose symmetry be obeyed with
respect to exchange of the pair of $Z$ bosons. This constraints the
number of independent helicity amplitudes to be less than or equal to
six. Even if the Spin-$J$ of $H$ is higher (i.e. $J\geqslant 3$), the
number of independent helicity amplitudes still remains six. However,
the helicity amplitudes corresponding to higher spin states involve
higher powers of momentum of $Z$, independent of the momentum
dependence of the form factors describing the process. We will show
that even for $J^P=2^+$ under a special case only two independent
helicity amplitudes may survive just as in the case of $J^P=0^+$. The
two cases are in principle indistinguishable unless one makes an
assumption on the momentum dependence of the form factors involved.

We start by considering the most general decay vertex for both scalar
and tensor resonances $H$ decaying to two $Z$ bosons. We evaluate the
partial decay rate of $H$ in terms of the invariant mass squared of
the dilepton produced from the non-resonant $Z$ and the angular
distributions of the four lepton final state. We demonstrate that by
studying three uni-angular distributions one can almost completely
determine the spin and parity of $H$ and also explore any anomalous
couplings in the most general fashion. We find that $J^{P}=0^{-}$ and
$2^-$ can easily be excluded.  The $J^{P}=0^{+}$ and $2^{+}$
possibilities can also be easily distinguished, but may require some
lepton invariant mass measurements if the most general tensor vertex
is considered.  Only if $H$ is found to be of Spin-2, a complete three
angle fit to the distribution is required to distinguish between
$J^{P}=2^{+}$ and $2^{-}$.

The determination of couplings and spin, parity of the boson is
important as there are other Spin-0 and Spin-2 particles predicted,
such as the $J=0$ radion \cite{Chacko:2012vm, Cheung:2011nv,
  deSandes:2011zs, Kubota:2012in, Grzadkowski:2012ng, Barger:2011hu,
  Rizzo:2002pq} and $J=2$ Kaluza-Klein graviton ~\cite{Han:1998sg,
  Fok:2012zk, Alves:2012fb, Geng:2012hy}, which can easily mimic the
initial signatures observed so far. Such cases have already been
considered in the literature even in the context of this
resonance. Our analysis is most general and such extensions are
limiting cases in our analysis as the couplings are defined by the
model.

In Section~\ref{sec:analysis} we layout the details of our analysis,
with Sections~\ref{subsec:scalar} and \ref{subsec:tensor} devoted
exclusively to Spin-0 Higgs and Spin-2 boson respectively.  A step by
step comparison with detailed procedure to distinguish the spin and
parity states of the new boson is discussed in
Sec.~\ref{subsec:comparison}.  In Sec.\ref{subsec:numerical} we
present a numerical study to demonstrate the discriminating power of
the uni-angular distribution analysis compared to the current
approach~\cite{Aad:2013xqa,Chatrchyan:2012jja}. We find that
uni-angular distribution is more powerful in discriminating between
the scalar ($0^+$) and pseudoscalar ($0^-$) hypothesis.

We conclude emphasizing the advantage of our approach in
Section~\ref{sec:conclusion}.

\begin{figure}[hbtp!]
\centering
\includegraphics{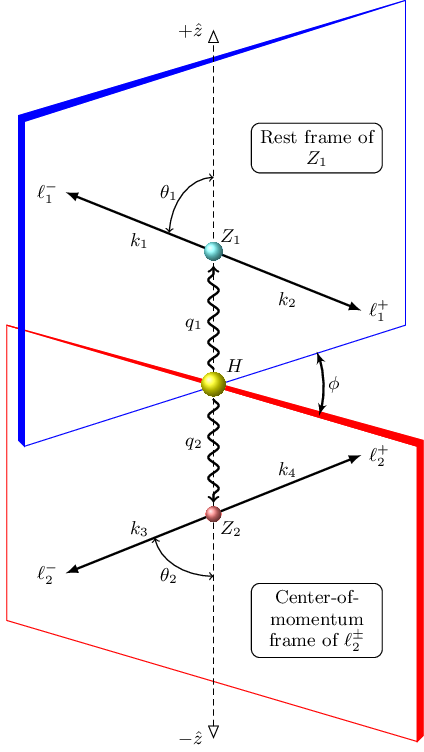}
\caption{Definition of the polar angles ($\theta_1$ and $\theta_2$)
  and the azimuthal angle ($\phi$) in the decay of Higgs ($H$) to a
  pair of $Z$'s, and then to four charged leptons: $H\to Z_1 + Z_2 \to
  (\ell_1^- + \ell_1^+) + (\ell_2^- + \ell_2^+),$ where $\ell_1,
  \ell_2 \in \{ e,\mu \}$. It should be clear from the figure that
  $\vec{k}_1 = -\vec{k}_2$ and $\vec{k}_3 = -\vec{k}_4$. Since $Z_2$
  is off-shell, we cannot go to its rest frame. However, given the
  momenta of $\ell_2^+$ and $\ell_2^-$ we can always go to their
  center-of-momentum frame.}
\label{fig:HZZst}
\end{figure}

\section{Decay of the new resonance to four charged leptons via two
  \texorpdfstring{$Z$}{Z} bosons}\label{sec:analysis}

Let us consider the decay of $H$ to four charged leptons via a pair of
$Z$ bosons: $$H \to Z_1 + Z_2 \to (\ell_1^- + \ell_1^+) + (\ell_2^- +
\ell_2^+),$$ where $\ell_1$, $\ell_2$ are leptons $e$ or $\mu$. As
mentioned in the introduction we assume $\ell_1$ and $\ell_2$ are not
identical. The kinematics for the decay is as shown in
Fig.~\ref{fig:HZZst}. The Higgs at rest is considered to decay with
the on-shell $Z_1$ moving along the $+\hat{z}$ axis and off-shell
$Z_2$ along the $-\hat{z}$ axis. The decays of $Z_1$ and $Z_2$ are
considered in their rest frame. The angles and momenta involved are as
described in Fig.~\ref{fig:HZZst}. The 4-momenta of $H$, $Z_1$ and
$Z_2$ are defined as $P$, $q_1$ and $q_2$ respectively. We choose
$Z_1$ to decay to lepton pair $\ell_1^\pm$ with momentum $k_1$ and
$k_2$ respectively and $Z_2$ to decay to $\ell_2^\pm$ with momentum
$k_3$ and $k_4$ respectively.

Nelson~\cite{Nelson:1984bb,Dell'Aquila:1985ve,Nelson:1986ki} and
Dell'Aquilla~\cite{Dell'Aquila:1985ve} realized the significance of
studying angular correlations in this process with Higgs boson
decaying to a pair of $Z$ bosons for inferring the nature of the Higgs
boson. Refs.~\cite{Miller:2001bi,Choi:2002dq, Choi:2002jk} were the
first to extend the analysis to include higher spin possibilities so
that any higher spin particle can effectively be distinguished from SM
Higgs. We study similar angular correlations in this paper. We begin
the study by considering the most general $HZZ$ vertices for a $J=0$
and a $J=2$ resonance $H$.  We shall first discuss the two spin
possibilities separately. Later we will layout the approach to
distinguish them assuming the most general $HZZ$ vertex.

\subsection{Spin-0 Higgs}\label{subsec:scalar}
The most general $HZZ$ vertex factor $V^{\alpha\beta}_{HZZ}$ for
Spin-0 Higgs is given by
\begin{equation}\label{eq:VHZZ0}
  V^{\alpha\beta}_{HZZ}=\displaystyle\frac{igM_Z}{\cos\theta_W} 
  \bigg( a \, g^{\alpha\beta} + b \, P^{\alpha}P^{\beta} 
  + i c \,\epsilon^{\alpha\beta\mu\nu} \; q_{1\mu}\,q_{2\nu} \bigg), 
\end{equation}
where $\theta_W$ is the \textit{weak mixing angle}, $g$ is the
electroweak coupling, and $a$, $b$, $c$ are some arbitrary form
factors dependent on the 4-momentum squares specifying the vertex. The
vertex $V^{\alpha\beta}_{HZZ}$ is derived from an effective Lagrangian
(see for example Ref.~\cite{Bolognesi:2012mm}) where higher
dimensional operators contribute to the momentum dependence of the
form factors. Since the effective Lagrangian in the case of arbitrary
new physics is not known, no momentum dependence of $a$, $b$ and $c$
can be assumed if the generality of the approach has to be
retained. Approaches using constant values for the form factors
therefore cannot provide unambiguous determination of spin-parity of
the new boson.  We emphasize that even though the momentum dependence
of $a$, $b$ and $c$ is not explicitly specified, they must be regarded
as being momentum dependent in general.  In SM, however, $a$, $b$, $c$
are constants and take the value $a=1$ and $b=c=0$ at tree level.

In Eq.~\eqref{eq:VHZZ0} the term proportional to $c$ is odd under
parity and the terms proportional to both $a$ and $b$ are even under
parity.  Partial-wave analysis tells that such a decay gets
contributions from the first three partial waves, namely
$\mathcal{S}$-wave, $\mathcal{P}$-wave and $\mathcal{D}$-wave.  Since
$\mathcal{S}$- and $\mathcal{D}$-waves are parity even while the
$\mathcal{P}$-wave is parity odd, the term associated with $c$
effectively describes the $\mathcal{P}$-wave contribution. The terms
proportional to $a$ and $b$ are admixtures of $\mathcal{S}$- and
$\mathcal{D}$-wave contributions. The decay of a Spin-0 particle to
two Spin-1 massive particles is hence always described by three
helicity amplitudes.

The decay under consideration is more conveniently described in terms
of helicity amplitudes $A_L$, $A_{\parallel}$ and $A_{\perp}$ defined
in the transversity basis as
\begin{align}
  A_L &= q_1 \cdot q_2 \, a+ M_H^2\, X^2 \, b, \label{eq:AL}\\
  A_{\parallel} &= \sqrt{2 q_1^2 \, q_2^2} \, a, \label{eq:AA}\\
  A_\perp &= \sqrt{2 q_1^2\, q_2^2}\, X \, M_H \, c, \label{eq:AP}
\end{align}
where $\sqrt{q_1^2}$ and $\sqrt{q_2^2}$ are the invariant masses of
the $\ell_1^{\pm}$ and $\ell_2^{\pm}$ lepton pairs, i.e.~$q_1^2 \equiv
(k_1+k_2)^2$, $q_2^2 \equiv (k_3+k_4)^2$,
\begin{equation}\label{eq:X}
  X=\dsp\frac{\sqrt{\dsp\lambda(M_H^2,q_1^2,q_2^{2})}}{\dsp 2M_H},
\end{equation}
$a$, $b$ and $c$ are the coefficients that enter the most general
vertex we have written in Eq.~\eqref{eq:VHZZ0} and
\begin{equation}
  \lambda(x,y,z)= x^2+y^2+z^2-2\,x\,y-2\,x\,z-2\,y\,z~.
\end{equation}
It should be remembered that the helicities $A_L, A_{\parallel}$ and
$A_{\perp}$ are in general functions of $q_1^2$ and $q_2^2$, even
though the functional dependence is not explicitly stated.  The
advantage of using the helicity amplitudes is that the helicity
amplitudes are orthogonal. Our helicity amplitudes are defined in the
transversity basis and thus differ from those given in
Ref.~\cite{Bolognesi:2012mm}. Our amplitudes can be classified by
their parity: $A_L$ and $A_\|$ are parity even and $A_\perp$ is parity
odd. This is unlike the amplitudes used in
Ref.~\cite{Bolognesi:2012mm}. Throughout the paper we use linear
combinations of the helicity amplitudes such that they have well
defined parity. {\em This basis may be referred to as the transversity
  basis}. Even though we work in terms of helicity amplitudes in the
transversity basis, we will show below, it is in fact possible to
uniquely extract out the coefficients $a,b,c$ which characterize the
most general $HZZ$ vertex for $J=0$ Higgs.

We will assume that $Z_1$ is on-shell while $Z_2$ is off-shell, unless
it is explicitly stated that both the $Z$ bosons are off-shell. The
off-shell nature of the $Z$ is denoted by a superscript `*'. One can
easily integrate over $q_1^2$ using the narrow width approximation of
the $Z$.  The helicity amplitudes are then defined at $q_1^2\equiv
M_Z^2$ and $q_2^2$.  In principle $q_1^2$ could also have been
explicitly integrated out in both the cases when either $Z_1$ is
off-shell or fully on-shell, resulting in some weighted averaged value
of the helicities. The differential decay rate for the process $H\to
Z_1 + Z_2^* \to (\ell_1^- + \ell_1^+) + (\ell_2^- + \ell_2^+)$, after
integrating over $q_1^2$ (assuming $Z_1$ is on-shell or even
otherwise) can now be written in terms of the angular distribution
using the vertex given in Eq.~\eqref{eq:VHZZ0} as:
\begin{widetext}
\begin{align}\label{eq:scalar-distribution}
  \frac{8\pi}{\gammaf}\frac{d^4\Gamma}{dq_2^2\; d\cos{\theta_1} \;
    d\cos{\theta_2} \; d\phi} &= 1 +\frac{|F_\||^2-|F_\perp|^2}{4}
  \cos\,2\phi\big(1-P_2(\cos\theta_1)\big)
  \big(1-P_2(\cos\theta_2)\big)\nn\\
  &\qquad +\frac{1}{2} \Im(F_{\parallel} F_{\perp}^*)\,\sin\,2\phi
  \big(1-P_2(\cos\theta_1)\big)\big(1-P_2(\cos\theta_2)\big)\nn\\
  &\qquad +\frac{1}{2}(1-3
  \modulus{F_L}^2)\,\big(P_2(\cos\theta_1)+P_2(\cos\theta_2)\big) +
  \frac{1}{4}(1+3\modulus{F_L}^2)\,P_2(\cos\theta_1)P_2(\cos\theta_2)\nn\\
  &\qquad +\frac{9}{8\sqrt{2}} \left( \Re(F_L
    F_{\parallel}^*)\,\cos\phi + \Im(F_L F_{\perp}^*)\,\sin\phi
  \right)
  \sin\,2\theta_1\,\sin\,2\theta_2\nn\\
  &\; +\eta\Bigg(\frac{3}{2} \Re(F_{\parallel}
  F_{\perp}^*)\big(\cos\theta_2 (2 + P_2(\cos\theta_1)) - \cos\theta_1
  (2 + P_2(\cos\theta_2))\big) \nn\\
  &\qquad \qquad +\frac{9}{2\sqrt{2}}\Re(F_L F_{\perp}^*)
  \big(\cos\theta_1-\cos\theta_2)\cos\phi \sin\theta_1\sin\theta_2\nn\\
  &\qquad \qquad -\frac{9}{2\sqrt{2}}\Im(F_L F_{\parallel}^*)
  \big(\cos\theta_1-\cos\theta_2)\sin\phi
  \sin\theta_1\sin\theta_2\Bigg)\nn\\
  &\quad
  -\frac{9}{4}\eta^2\Bigg((1-\modulus{F_L}^2)\cos\theta_1\cos\theta_2
  + \sqrt{2} \left( \Re(F_L F_{\parallel}^*)\cos\phi + \Im(F_L
    F_{\perp}^*)\sin\phi \right) \sin\theta_1\sin\theta_2\Bigg),
\end{align}
\end{widetext}
where the {\it{helicity fractions}} $F_L$, $F_{\parallel}$ and
$F_{\perp}$ are defined as
\begin{equation}
  \dsp F_\lambda=\frac{A_\lambda}{\sqrt{ \modulus{A_L}^2 +
      \modulus{A_{\parallel}}^2 + \modulus{A_{\perp}}^2}}, 
\end{equation}
where $\lambda \in \{ L, \parallel, \perp \}$ and
\begin{align}
  \gammaf & \equiv \frac{d \Gamma}{dq_2^2} = \mathcal{N} \left(
    \modulus{A_L}^2 + \modulus{A_{\parallel}}^2 +
    \modulus{A_{\perp}}^2 \right) \label{eq:gamma},\\%
  \text{with } \mathcal{N} &= \frac{1}{2^{4}} \; \frac{1}{\pi^2} \;
  \frac{g^2}{\cos^2{\theta_W}} \; \frac{\text{Br}_{\ell\ell}^2}{M_H^2}
  \; \frac{\Gamma_Z}{M_Z} \nn\\%
  & \quad \times \frac{X}{\left( \left( q_2^2 - M_Z^2 \right)^2 +
      M_Z^2 \Gamma_Z^2 \right)}.%
\end{align}
where $\Gamma_Z$ is the total decay width of the $Z$ boson,
$\text{Br}_{\ell\ell}$ is the branching ratio for the decay of $Z$
boson to two mass-less leptons: $Z \to \ell^+ \ell^-$ and we have used
the narrow width approximation for the on-shell $Z$.  We emphasize
that with $q_1^2$ integrated out the helicity amplitudes $A_\lambda$
and helicity fractions $F_\lambda$ are functions only of $q_2^2$.  In
Eq.~\eqref{eq:scalar-distribution} $\eta$ is defined as
\begin{equation}
  \eta=\frac{2 v_{\sss\ell} a_{\sss\ell}}{v_{\sss\ell}^2+a_{\sss\ell}^2}
\end{equation}
with $v_{\sss\ell}=2 I_{3\ell}-4e_{\sss\ell} \sin^2\theta_W$ and
$a_{\sss\ell}=2 I_{3\ell}$, and $P_2(x)$ is the 2nd degree Legendre
polynomial:
\begin{equation}
  P_2(x) = \frac{1}{2}(3x^2-1) \qquad \mbox{(with $x \in
    \{ \cos\theta_1, \cos\theta_2 \}$)}.
\end{equation}

We have chosen to express the the differential decay rate in terms of
Legendre polynomials for $\cos{\theta_1}$ and $\cos{\theta_2}$ and
Fourier series for $\phi$. This ensures that each term in
Eq.~\eqref{eq:scalar-distribution} is orthogonal to any other term in
the distribution.  The Legendre polynomials $P_m(\cos{\theta_1})$ and
$P_m(\cos{\theta_2})$ satisfy the orthogonality condition since the
range of $\cos{\theta_1}$ and $\cos{\theta_2}$ is $-1$ to $1$, whereas
that of $\phi$ is $0$ to $2\,\pi$.  Our approach of using Legendre
polynomials and the choice of helicity amplitudes in transversity
basis classified by parity form the corner-stone of our analysis. The
same technique will be used in Sec.~\ref{subsec:tensor} to analyze the
Spin-2 case.

An interesting observation in the scalar case is that the coefficients
of $P_2 (\cos\theta_1)$ and $P_2 (\cos\theta_2)$ are identically equal
to $\frac{1}{2}(1-3 |F_L|^2)$ in both magnitude and sign. It is worth
noting that the coefficients of $\cos2\phi \; P_2 (\cos\theta_1)$ and
$\cos2\phi \; P_2 (\cos\theta_2)$ are also identically equal to
$\frac{1}{4}(|F_{\parallel}|^2-|F_\perp|^2)$ in both magnitude and
sign.

Integrating Eq.~\eqref{eq:scalar-distribution} with respect to
$\cos{\theta_1}$ or $\cos{\theta_2}$ or $\phi$, the following
uni-angular distributions are obtained:
\begin{widetext}
\begin{align}
  \frac{1}{\gammaf}\frac{d^2\Gamma}{dq_2^2 \; d\cos\theta_1} &=
  \frac{1}{2} + T_2^{(0)}\,P_2(\cos\theta_1) - T_1^{(0)}
  \cos\theta_1, \label{eq:scalar-costheta1} \\
  \frac{1}{\gammaf}\frac{d^2\Gamma}{dq_2^2 \; d\cos\theta_2} &=
  \frac{1}{2} + T_2^{(0)}\,P_2(\cos\theta_2) + T_1^{(0)}
  \cos\theta_2, \label{eq:scalar-costheta2} \\
  \frac{2\pi}{\gammaf}\frac{d^2\Gamma}{dq_2^2 \; d\phi} &= 1 +
  U_2^{(0)}\,\cos\,2\phi + V_2^{(0)}\,\sin\,2\phi + U_1^{(0)}\cos\phi
  + V_1^{(0)} \sin\phi, \label{eq:scalar-phi}
\end{align}
\end{widetext}
where
\begin{align}
\label{eq:T10}
T_2^{(0)} &=\frac{1}{4} (1-3\modulus{F_L}^2),\\
\label{eq:T20}
U_2^{(0)} &=\frac{1}{4} (|F_{\parallel}|^2-|F_{\perp}|^2),\\
\label{eq:T30}
V_2^{(0)} &=\frac{1}{2}\,\mathrm{Im}(F_{\parallel}F_{\perp}^*),\\
T_1^{(0)} &=\frac{3}{2} \, \eta \Re(F_{\parallel} F_{\perp}^*),\\
U_1^{(0)} &=-\frac{9\pi^2}{32\sqrt{2}} \eta^2 \, \Re(F_L F_{\parallel}^*),\\
V_1^{(0)} &=-\frac{9\pi^2}{32\sqrt{2}} \eta^2 \, \Im(F_L F_{\perp}^*),
\end{align}
are explicitly functions of $q_2^2$. The superscript $(0)$ indicates
the spin of $H$. Since $P_0(\cos \theta_{1,2})=1$, $P_1(\cos
\theta_{1,2})= \cos\theta_{1,2}$, $P_2(\cos \theta_1)$, $\cos\phi$,
$\sin\phi$, $\cos 2\phi$ and $\sin 2\phi$ are orthogonal functions,
the coefficients of each of the terms can be extracted individually.
We can also extract all the above coefficients in terms of asymmetries
defined as below:
\begin{widetext}
\begin{align}
  T_1^{(0)} &= \left( \int_{-1}^{0} - \int_{0}^{+1} \right)
  d\cos\theta_1 \; \left( \frac{1}{\gammaf}\frac{d^2\Gamma}{dq_2^2 \;
    d\cos\theta_1} \right) = \left( -\int_{-1}^{0} + \int_{0}^{+1}
  \right) d\cos\theta_2 \; \left( \frac{1}{\gammaf}
  \frac{d^2\Gamma}{dq_2^2 \; d\cos\theta_2} \right), \label{eq:T10asym}\\%
  T_2^{(0)} &= \frac{4}{3} \left( \int_{-1}^{-\frac{1}{2}} -
  \int_{-\frac{1}{2}}^{+\frac{1}{2}} + \int_{+\frac{1}{2}}^{+1}
  \right) d\cos\theta_{1,2} \; \left( \frac{1}{\gammaf}
  \frac{d^2\Gamma}{dq_2^2 \; d\cos\theta_{1,2}} \right),\label{eq:T20asym}\\%
  U_1^{(0)} &= \frac{1}{4} \left( -\int_{-\pi}^{-\frac{\pi}{2}} +
  \int_{-\frac{\pi}{2}}^{+\frac{\pi}{2}} -
  \int_{+\frac{\pi}{2}}^{+\pi} \right) d\phi \; \left(
  \frac{2\pi}{\gammaf} \frac{d^2\Gamma}{dq_2^2 \; d\phi}
  \right), \label{eq:U10asym}\\%
  U_2^{(0)} &= \frac{1}{4} \left( \int_{-\pi}^{-\frac{3\pi}{4}} -
  \int_{-\frac{3\pi}{4}}^{-\frac{\pi}{4}} +
  \int_{-\frac{\pi}{4}}^{\frac{\pi}{4}}
  -\int_{\frac{\pi}{4}}^{\frac{3\pi}{4}} + \int_{\frac{3\pi}{4}}^{\pi}
  \right) d\phi \; \left( \frac{2\pi}{\gammaf}\frac{d^2\Gamma}{dq_2^2
    \; d\phi} \right),\label{eq:U20asym}\\%
  V_1^{(0)} &= \frac{1}{4} \left( - \int_{-\pi}^{0} + \int_{0}^{+\pi}
  \right) d\phi \; \left( \frac{2\pi}{\gammaf}\frac{d^2\Gamma}{dq_2^2
    \; d\phi} \right),\label{eq:V10asym}\\%
  V_2^{(0)} &= \frac{1}{4} \left( \int_{-\pi}^{-\frac{\pi}{2}} -
  \int_{-\frac{\pi}{2}}^{0} + \int_{0}^{+\frac{\pi}{2}} -
  \int_{+\frac{\pi}{2}}^{+\pi} \right) d\phi \; \left(
  \frac{2\pi}{\gammaf}\frac{d^2\Gamma}{dq_2^2 \; d\phi}
  \right).\label{eq:V20asym}%
\end{align}
\end{widetext}
As had already been realized from Eq.~\eqref{eq:scalar-distribution},
the coefficients of $P_2(\cos\theta_1)$ and $P_2(\cos\theta_2)$ as
well as the coefficients of $\cos\theta_1$ and $\cos\theta_2$ in
Eqs.~\eqref{eq:scalar-costheta1} and \eqref{eq:scalar-costheta2} are
identical. This results in a maximum of 6 possible independent
measurements $T_1^{(0)}$, $U_1^{(0)}$, $V_1^{(0)}$, $T_2^{(0)}$,
$U_2^{(0)}$ and $V_2^{(0)}$ using uni-angular analysis. For the decay
under consideration, $v_{\sss\ell} = -1 + 4\sin^2{\theta_W}$ and
$a_{\sss\ell} = -1$. Substituting the experimental value for the weak
mixing angle: $\sin^2{\theta_W} = 0.231$, we get $\eta = 0.151$ and
$\eta^2 = 0.0228$. Owing to such small values of $\eta$ and $\eta^2$
it is unlikely that $T_1^{(0)}$, $U_1^{(0)}$ and $V_1^{(0)}$ can be
measured using the small data sample current available at LHC,
reducing the number of independent measurable to three.

Using Eqs.~\eqref{eq:T10} and \eqref {eq:T20} and the identity
$\modulus{F_L}^2 + \modulus{F_{\parallel}}^2 + \modulus{F_{\perp}}^2 =
1$, the following solutions for $\modulus{F_L}^2$,
$\modulus{F_{\parallel}}^2$ and $\modulus{F_{\perp}}^2$ are obtained:
\begin{align}
  \modulus{F_L}^2 &= \frac{1}{3} \left( 1 - 4\,T_2^{(0)} \right),\label{eq:FL} \\
  \modulus{F_{\parallel}}^2 &= \frac{1}{3} \left( 1 + 2\, T_2^{(0)}
  \right) + 2\, U_2^{(0)}, \label{eq:Fparallel}\\
  \modulus{F_{\perp}}^2 &= \frac{1}{3} \left( 1 + 2\, T_2^{(0)}
  \right) - 2\, U_2^{(0)}.\label{eq:Fperp}
\end{align}

\begin{figure}
\centering
  \subfloat[Plot of $T_2^{(0)}$ and $U_2^{(0)}$ vs $\sqrt{q_2^2}$.]{\label{fig:T1T2}
    \includegraphics[scale=0.815]{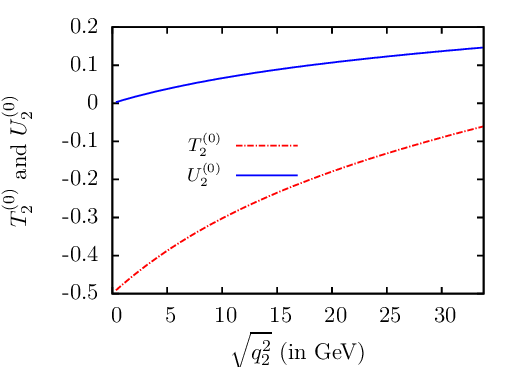}}\\%
  \subfloat[Plot of $F_L$ and $F_{\parallel}$
  vs. $\sqrt{q_2^2}$.]{\label{fig:FLFA}
    \includegraphics[scale=0.815]{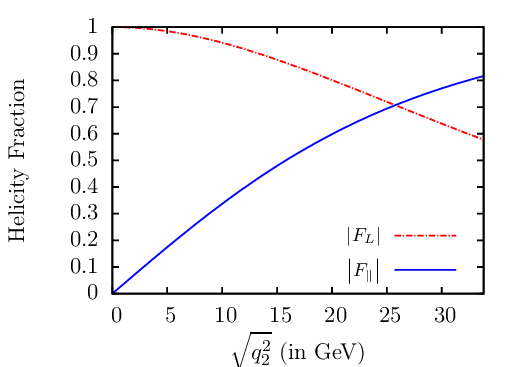}}\\%
  \subfloat[Plot of $\displaystyle
  \frac{1}{\Gamma}\frac{d\Gamma}{dq_2^2}$ vs $\sqrt{q_2^2}$.]{\label{fig:Gama-q}
    \includegraphics[scale=0.815]{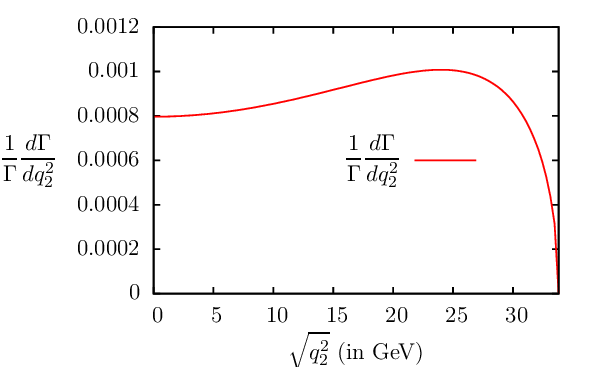}}
  \caption{Plots of various observables in SM only. We have used $M_H
    = 125 \; \gev$, $\sqrt{q_1^2}= 91.18 \; \gev$ for the above
    plots. The integrated values for the observables $T_2^{(0)}$ and
    $U_2^{(0)}$ are uniquely predicted in SM at tree level to be
    $-0.148$ and $0.117$ respectively. }
\label{fig:SM-Higgs}
\end{figure}

We have shown that one can easily measure all the three helicity
fractions using uni-angular distributions.  We can also measure
$\mathrm{Im}(F_{\parallel}F_{\perp}^*)$, which is proportional to sine
of the phase difference between the two helicity amplitudes
$A_{\parallel}$ and $A_{\perp}$. In other words, we can also measure
the relative phase between the parity-odd and parity-even
amplitudes. Such a phase can arise if $CP$-symmetry is violated in
$HZZ$ interactions or could indicate pseudo-time reversal violation
arising from loop level contributions or rescattering effects akin to
the strong phase in strong interactions. Since such a term requires
contributions from both parity-even and parity-odd partial waves,
$V_2^{(0)}=0$ in SM.  In the case of SM we have $a=1$ and
$b=c=0$. Assuming narrow width approximation for the on-shell $Z_1$ we
get
\begin{align}
  F_{\perp} &= 0, \\
  \frac{F_L}{F_{\parallel}} &\equiv \mathsf{T}= \frac{M_H^2 - M_Z^2 -
    q_2^2}{2\sqrt{2} M_Z \sqrt{q_2^2}}.
\end{align}
Clearly, for the case of SM the term $\mathsf{T}$ has a characteristic
dependence on $\sqrt{q_2^2}$.  Demanding $F_{\perp} = 0$, we get
\begin{equation}
  \label{eq:T2-scalar}
  U_2^{(0)} = \frac{1}{6} \left( 1 + 2  \, T_2^{(0)} \right),
\end{equation}
and
\begin{equation}
  \modulus{\mathsf{T}} = \frac{1-4\,T_2^{(0)}}{2+4\,T_2^{(0)}}.
\end{equation}
Thus for SM we can predict the experimental values for the
coefficients $T_2^{(0)}$ and $U_2^{(0)}$ as:
\begin{equation}\label{eq:T2U2-SMHiggs}
  T_2^{(0)} = \frac{1}{4} \left(
    \frac{1-2\modulus{\mathsf{T}}}{1+\modulus{\mathsf{T}}} \right), 
  \qquad U_2^{(0)} = \frac{1}{4 \left( 1+\modulus{\mathsf{T}} \right)}.
\end{equation}
It is evident that $T_2^{(0)}$ and $U_2^{(0)}$ are functions of
$\sqrt{q_2^2}$ alone and are uniquely predicted in the SM.
$T_2^{(0)}$ and $U_2^{(0)}$ are pure numbers for a given value of
$\sqrt{q_2^2}$.  Their variation with respect to $\sqrt{q_2^2}$ is
shown in Fig.~\ref{fig:T1T2}.  It is clear from the plot that
$T_2^{(0)}$ is always negative while $U_2^{(0)}$ is always positive in
the SM.  The variation of the helicity fractions with respect to
$\sqrt{q_2^2}$ is shown in Fig.~\ref{fig:FLFA}. Fig.~\ref{fig:Gama-q}
also shows the variation of the normalized differential decay width of
the SM Higgs decaying to four charged leptons via two $Z$ bosons, with
respect to $\sqrt{q_2^2}$. Fig.~\ref{fig:SM-Higgs} contains all the
vital experimental signatures of the SM Higgs and must be verified in
order for the new boson to be consistent with the SM Higgs boson. We
emphasize that a nonzero measurement of $F_{\perp}$ will be a litmus
test indicating a non-SM behavior for the Higgs. Furthermore, a
non-zero $V_2^{(0)}$ would imply that the observed resonance is not of
definite parity.

If we find the new boson to be of $J^{PC} = 0^{++}$, but still not
exactly like the SM Higgs, then we need to know the values of $a$ and
$b$ in the vertex factor of Eq.~\eqref{eq:VHZZ0}. It is easy to find
that for a general $0^{++}$ boson, the values of both $a$ and $b$ are
given by
\begin{align}
  a &= \frac{F_{\parallel} \sqrt{\gammaf/\mathcal{N}}}{\sqrt{2} M_Z
    \sqrt{q_2^2}}, \label{eq:a}\\
  b &= \frac{\sqrt{\gammaf/\mathcal{N}}}{M_H^2 X^2}\left( F_L -
    \frac{M_H^2 - M_Z^2 - q_2^2}{2\sqrt{2} M_Z \sqrt{q_2^2} } F_{\parallel}
  \right). \label{eq:b}
\end{align}
For SM $a=1$ and $b=0$ at tree level only. At loop level even within
SM these values would differ. It may be hoped that $a$ and $b$
determined in this way may enable testing SM even at one loop level
once sufficient data is acquired. This is significant as triple-Higgs
vertex contributes at one loop level and {\it measurement of $b$ may
  provide the first verification of the Higgs-self coupling}. Even if
the scalar boson is not a parity eigenstate but an admixture of even
and odd parity states, Eqs.~\eqref{eq:a} and \eqref{eq:b} can be used
to determine $a$ and $b$.  We can determine $c$ by measuring
$F_{\perp}$:
\begin{equation}\label{eq:c}
  c = \frac{F_{\perp} \sqrt{\gammaf/\mathcal{N}}}{\sqrt{2} M_Z \sqrt{q_2^2} M_H X},
\end{equation}
Therefore, it is possible to get exact solutions for $a,b,c$ in terms
of the experimentally observable quantities like $F_L$,
$F_{\parallel}$, $F_{\perp}$ and $\gammaf$.

We want to stress that it is impossible to extract out both $a$ and
$b$ by measuring only one uni-angular distribution (corresponding to
either $\cos{\theta_1}$ or $\cos{\theta_2}$), since the helicity
amplitude $A_L$ contains both $a$ and $b$.  Hence, it is not possible
to conclude that the $0^{++}$ boson is a Standard Model Higgs by
studying $\cos{\theta_1}$ or $\cos{\theta_2}$ distributions alone.


The current data set is limited and may allow binning only in one
variable. We therefore examine what conclusions can be made if $q_2^2$
is also integrated out and only the three uni-angular distributions
are studied individually. As can be seen from Eqs.~\eqref{eq:a},
\eqref{eq:b} and \eqref{eq:c} we can obtain some weighted averages of
$a$ and $c$. These equations will only allow us to verify whether
$a=1$ and $c=0$. In addition the presence of any phase between the
parity-even and parity-odd amplitudes can still be inferred from
Eq.~\eqref{eq:T30}. The integrated values for the observables
$T_2^{(0)}$ and $U_2^{(0)}$ are uniquely predicted in SM at tree level
to be $-0.148$ and $0.117$ respectively.

\subsection{Spin-2 Boson}\label{subsec:tensor}
As stated in the Introduction we shall use the same symbol $H$ to
denote the boson even if it is of Spin-2.  The most general $HZZ$
vertex factor $V^{\mu\nu;\alpha\beta}_{HZZ}$ for Spin-2 boson, with
polarization $\epsilon^{\mu\nu}_{\sss(T)}$ has the following tensor
structure
\begin{widetext}
\begin{eqnarray}
  V^{\mu\nu;\alpha\beta}_{HZZ}&=& A \left(g^{\alpha\nu} \;
    g^{\beta\mu} + g^{\alpha\mu} \; g^{\beta\nu}\right) + B \left(
    Q^{\mu} \left(Q^{\alpha} \; g^{\beta\nu} + Q^{\beta} \;
      g^{\alpha\nu} \right)
    + Q^{\nu} \left(Q^{\alpha} \; g^{\beta\mu} + Q^{\beta} \;
      g^{\alpha\mu}\right) \right) + C
  \left( Q^{\mu} \; Q^{\nu} \; g^{\alpha\beta} \right) \nn\\
  &&- D \left( Q^{\alpha} \; Q^{\beta} \; Q^{\mu} \; Q^{\nu} \right)
  +2 i \; E \big( g^{\beta\nu} \; \epsilon^{\alpha\mu\rho\sigma}
  - g^{\alpha\nu} \; \epsilon^{\beta\mu\rho\sigma} + g^{\beta\mu} \;
  \epsilon^{\alpha\nu\rho\sigma}
  - g^{\alpha\mu} \; \epsilon^{\beta\nu\rho\sigma} \big) q_{1\rho}
  q_{2\sigma} \nn \\
  &&+ i \, F \big( Q^{\beta} \left(Q^{\nu} \, \epsilon^{\alpha\mu\rho\sigma}
    + Q^{\mu} \, \epsilon^{\alpha\nu\rho\sigma}\right) - Q^{\alpha}
  \left(Q^{\nu} \, \epsilon^{\beta\mu\rho\sigma}
    + Q^{\mu} \, \epsilon^{\beta\nu\rho\sigma}\right) \big) q_{1\rho} 
  q_{2\sigma}, \label{eq:VHZZ2}
\end{eqnarray}
\end{widetext}
where $\epsilon_\alpha$ and $\epsilon_\beta$ are the polarizations of
the two $Z$ bosons; $A$, $B$, $C$, $D$, $E$ and $F$ are arbitrary
coefficients and $Q$ is the difference of the four momenta of the two
$Z$'s, i.e. $Q=q_1-q_2$. Only the term that is associated with the
coefficient $A$ is dimensionless.  The form of the vertex factor
ensures that $P_{\mu} \epsilon^{\mu\nu}_{\sss(T)} = P_{\nu}
\epsilon^{\mu\nu}_{\sss(T)} = 0$ and $g_{\mu\nu}
\epsilon^{\mu\nu}_{\sss(T)} = 0$, which stem from the fact that the
field of a Spin-2 particle is described by a symmetric, traceless
tensor with null four-divergence. Here like the Spin-0 case $P$ is the
sum of the four-momenta of the two $Z$'s, i.e. $P = q_1 + q_2 $. Since
we are considering the decay of Higgs to two $Z$ bosons, the vertex
factor must be symmetric under exchange of the two identical
bosons. This is taken care of by making the vertex factor symmetric
under simultaneous exchange of $\alpha,\beta$ and corresponding
momenta of $Z_1$ and $Z_2$.  The Lagrangian that gives rise to the
vertex factor $V^{\mu\nu;\alpha\beta}_{HZZ}$ contains higher
dimensional operators, which are responsible for the momentum
dependence of the form factors.

In $V^{\mu\nu;\alpha\beta}_{HZZ}$ the terms that are proportional to
$E$ and $F$ are parity-odd and the rest of the terms in
$V^{\mu\nu;\alpha\beta}_{HZZ}$ are parity-even. From helicity analysis
it is known that the decay of a massive Spin-2 particle to two
identical, massive, Spin-1 particles is described by six helicity
amplitudes. Bose symmetry between the pair of Z bosons
\cite{Jacob:1959at,Chung:1102240} imposes constraints on the vertex
$V^{\mu\nu;\alpha\beta}_{HZZ}$ such that it gets contributions from
two parity-odd terms that are admixture of one $\mathcal{P}$-wave and
one $\mathcal{F}$-wave, and four parity-even terms that are some
combinations of one $\mathcal{S}$-wave, two $\mathcal{D}$-waves and
one $\mathcal{G}$-wave contributions. Even for the case of Spin-2
boson we choose to work with helicity amplitudes as they are
orthogonal but choose a basis such that amplitudes have definite
parity associated with them. We find the following six helicity
amplitudes in transversity basis:
\begin{align}
  A_L &= \frac{4 X}{3 \mathsf{u}_1} \bigg( E \left( \mathsf{u}_2^4 -
    M_H^2 \mathsf{u}_1^2 \right) + F \left( 4 \mathsf{u}_1^2 M_H^2 X^2 \right) \bigg),\\
  A_M &= \frac{8 \sqrt{q_1^2 \, q_2^2} \mathsf{v} X}{3 \sqrt{3} \mathsf{u}_1} \; E ,\\
  A_1 &= \frac{2 \sqrt{2}}{3 \sqrt{3} M_H^2} \bigg( A \left( M_H^4 -
    \mathsf{u}_2^4 \right) - B \left( 8 M_H^4 X^2 \right) \nonumber \\
  & \qquad \qquad \qquad + C \left( 4M_H^2 X^2 \right) \left(
    \mathsf{u}_1^2 - M_H^2 \right) \nonumber \\
  & \qquad \qquad \qquad \qquad \qquad \qquad - D \left( 8 M_H^4 X^4
  \right) \bigg), \label{eq:A1}\\
  A_2 &= \frac{8 \sqrt{q_1^2 \, q_2^2}}{3 \sqrt{3}} \left( A + 4 X^2 C \right),\label{eq:A2}\\
  A_3 &= \frac{4}{3 M_H \mathsf{u}_1} \bigg( A \left( \mathsf{u}_2^4 -
    M_H^2 \mathsf{u}_1^2 \right) + B \left( 4 \mathsf{u}_1^2 M_H^2 X^2
  \right) \bigg),\\
  A_4 &= \frac{8 \sqrt{q_1^2 \, q_2^2} \mathsf{w}}{3 M_H \mathsf{u}_1} \; A,
\end{align}
where $\mathsf{u}_1$, $\mathsf{u}_2$, $\mathsf{v}$ and $\mathsf{w}$
are defined as
\begin{align}
  \mathsf{u}_1^2 &= q_1^2 + q_2^2,\\
  \mathsf{u}_2^2 &= q_1^2 - q_2^2,\\
  \mathsf{v}^2 &= 4 M_H^2 \mathsf{u}_1^2 + 3 \mathsf{u}_2^4,\\
  \mathsf{w}^2 &= 2 M_H^2 \mathsf{u}_1^2 + \mathsf{u}_2^4.
\end{align}
The quantity $X$ is as defined in Eq.~\eqref{eq:X}.

We wish to clarify that our vertex factor
$V^{\mu\nu;\alpha\beta}_{HZZ}$ is the most general one. An astute
reader can easily write down terms that are not included in our vertex
and wonder how such a conclusion of generality can be made. For
example, one can add a new possible term such as $i\,G\,\left(
\epsilon^{\alpha\beta\nu\rho} P_{\rho} Q^{\mu} +
\epsilon^{\alpha\beta\mu\rho} P_{\rho} Q^{\nu} \right)$.  It is easy
to verify that this new form factor $G$ enters our helicity amplitudes
$A_L$ and $A_M$ in the combination $(E-2G)$:
\begin{align}
  A_L &= \frac{4 X}{3 \mathsf{u}_1} \bigg( \left(E-2G\right) \left(
    \mathsf{u}_2^4 -
    M_H^2 \mathsf{u}_1^2 \right) + F \left( 4 \mathsf{u}_1^2 M_H^2 X^2 \right) \bigg),\\
  A_M &= \frac{8 \sqrt{q_1^2 \, q_2^2} \mathsf{v} X}{3 \sqrt{3} \mathsf{u}_1} \;
  \left(E-2G\right).
\end{align}
Note that only this combination of $E$ and $G$ is accessible to
experiments and all other helicity amplitudes remain unchanged.
Since, there exist only six independent helicity amplitudes
corresponding to six partial waves for the Spin-2 case, the number of
helicity amplitudes in the transversity basis must also be six.
Adding any new terms to the vertex factor will simply modify the
expressions for the helicity amplitudes. The generality of our vertex
$V^{\mu\nu;\alpha\beta}_{HZZ}$ is therefore very robust. Having
established the generality of $V^{\mu\nu;\alpha\beta}_{HZZ}$ we will
henceforth not consider any term absent in the vertex of
Eq.~\eqref{eq:VHZZ2}. Our helicity amplitudes are different from those
given in Ref.~\cite{Bolognesi:2012mm}.  In
Ref.~\cite{Bolognesi:2012mm}, they provide eight independent helicity
amplitudes.  If we consider the Bose symmetry of the two identical
vector bosons to which $H$ is decaying, then these should reduce to
six independent helicity amplitudes.  Again as stated in the scalar
case, our helicity amplitudes are classified by their parity and thus
differ from those in Ref.~\cite{Bolognesi:2012mm}. Our amplitudes
$A_L$ and $A_M$ have parity-odd behavior, and the rest of the helicity
amplitudes have parity-even behavior. In contrast not all the
amplitudes enunciated in Ref.~\cite{Bolognesi:2012mm} have clear
parity characteristics.

Once again just as in the scalar case we will start by assuming that
$Z_1$ is on-shell while $Z_2$ is off-shell. The integration over
$q_1^2$ is done using the narrow width approximation of the $Z$. In
tensor case, however, off-shell $Z_1$ will also have to be considered
in a special case.  We hence consider that $q_1^2$ is explicitly
integrated out whether $Z_1$ is off-shell or fully on-shell.  In case
$Z_1$ is off-shell the resulting helicities are some weighted averaged
value and should not be confused with well defined values at
$q_1^2\equiv M_Z^2$.  The differential decay rate for the process
$H\to Z_1 + Z_2^* \to (\ell_1^- + \ell_1^+) + (\ell_2^- + \ell_2^+)$,
after integrating over $q_1^2$ (assuming $Z_1$ is on-shell or even
otherwise) can now be written in terms of the angular distribution
using the vertex given in Eq.~\eqref{eq:VHZZ2} as:
\begin{widetext}
  \begin{align}
    & \frac{8 \pi}{\gammaf} \frac{d^4 \Gamma}{dq_2^2 \;
      d\cos{\theta_1} \;
      d\cos{\theta_2} \; d\phi} \nonumber \\
    &= 1 + \left( \frac{1}{4} \modulus{F_2}^2 - \left( M_H^2
        \frac{\mathsf{u}_1^2}{\mathsf{v}^2} \right) \modulus{F_M}^2
    \right) \; \cos{2\phi} \; (1-P_2(\cos{\theta_1})) \;
    (1-P_2(\cos{\theta_2})) \nonumber \\
    & \quad + \left( M_H \frac{\mathsf{u}_1}{\mathsf{v}} \right)
    \Im(F_2 F_M^*) \; \sin{2\phi} \; (1-P_2(\cos{\theta_1})) \;
    (1-P_2(\cos{\theta_2})) \nonumber \\
    & \quad + \frac{P_2(\cos{\theta_1})}{2} \Bigg( \left(-2
      \modulus{F_1}^2 + \modulus{F_2}^2 \right) + \left(
      \modulus{F_3}^2+\modulus{F_L}^2 \right) \left( \frac{q_1^2-2
        q_2^2}{\mathsf{u}_1^2} \right) \nonumber \\
    & \qquad \qquad \qquad \qquad + \modulus{F_M}^2 \left( 4 M_H^2
      \frac{\mathsf{u}_1^2}{\mathsf{v}^2} +3
      \frac{\mathsf{u}_2^4}{\mathsf{u}_1^2 \mathsf{v}^2} \left(q_2^2-2
        q_1^2\right) \right) + \modulus{F_4}^2 \left( 2 M_H^2
      \frac{\mathsf{u}_1^2}{\mathsf{w}^2} +
      \frac{\mathsf{u}_2^4}{\mathsf{u}_1^2 \mathsf{w}^2} \left(q_2^2 -
        2
        q_1^2\right) \right) \nonumber \\
    & \qquad \qquad \qquad \qquad + \left( 6 \sqrt{q_1^2 \, q_2^2}
      \frac{\mathsf{u}_2^2}{\mathsf{u}_1^2 \mathsf{w}} \right) \;
    \Re(F_3 F_4^*) + \left( 6 \sqrt{3} \sqrt{q_1^2 \, q_2^2}
      \frac{\mathsf{u}_2^2}{\mathsf{u}_1^2 \mathsf{v}} \right) \;
    \Re(F_L
    F_M^*) \Bigg) \nonumber \\
    & \quad + \frac{P_2(\cos{\theta_2})}{2} \Bigg( \left( -2
      \modulus{F_1}^2 + \modulus{F_2}^2 \right) + \left(
      \modulus{F_3}^2+\modulus{F_L}^2 \right) \left( \frac{q_2^2 - 2
        q_1^2}{\mathsf{u}_1^2} \right) \nonumber \\
    & \qquad \qquad \qquad \qquad + \modulus{F_M}^2 \left( 4 M_H^2
      \frac{\mathsf{u}_1^2}{\mathsf{v}^2} + 3
      \frac{\mathsf{u}_2^4}{\mathsf{u}_1^2 \mathsf{v}^2} \left(q_1^2-2
        q_2^2 \right) \right) +\modulus{F_4}^2 \left( 2 M_H^2
      \frac{\mathsf{u}_1^2}{\mathsf{w}^2} +
      \frac{\mathsf{u}_2^4}{\mathsf{u}_1^2 \mathsf{w}^2} \left(q_1^2-2
        q_2^2\right) \right) \nonumber \\
    & \qquad \qquad \qquad \qquad -\left( 6 \sqrt{q_1^2 \, q_2^2}
      \frac{\mathsf{u}_2^2}{\mathsf{u}_1^2 \mathsf{w}} \right) \;
    \Re(F_3 F_4^*) - \left( 6\sqrt{3} \sqrt{q_1^2 \, q_2^2}
      \frac{\mathsf{u}_2^2}{\mathsf{u}_1^2
        \mathsf{v}} \right) \; \Re(F_L F_M^*) \Bigg) \nonumber \\
    & \quad +\frac{P_2(\cos{\theta_1}) P_2(\cos{\theta_2})}{2} \Bigg(
    2 \modulus{F_1}^2 + \frac{1}{2} \modulus{F_2}^2 - \modulus{F_3}^2
    - \modulus{F_L}^2 - \left( \frac{\mathsf{u}_2^4-M_H^2
        \mathsf{u}_1^2}{\mathsf{w}^2} \right) \modulus{F_4}^2 + \left(
      \frac{2 M_H^2 \mathsf{u}_1^2-3 \mathsf{u}_2^4}{\mathsf{v}^2}
    \right)
    \modulus{F_M}^2 \Bigg) \nonumber \\
    & \quad +\frac{9 \sin{2\theta_1} \sin{2\theta_2} \cos{\phi}}{16}
    \Bigg( \left( \modulus{F_3}^2 - \modulus{F_L}^2 \right) \left(
      \frac{\sqrt{q_1^2 \, q_2^2}}{\mathsf{u}_1^2} \right) + 3 \modulus{F_M}^2
    \left( \sqrt{q_1^2 \, q_2^2} \frac{\mathsf{u}_2^4}{\mathsf{u}_1^2 \mathsf{v}^2}
    \right) - \modulus{F_4}^2 \left( \sqrt{q_1^2 \, q_2^2}
      \frac{\mathsf{u}_2^4}{\mathsf{u}_1^2
        \mathsf{w}^2} \right) \nonumber \\
    & \qquad \qquad \qquad \qquad \qquad \qquad - \left(
      \frac{\mathsf{u}_2^4}{\mathsf{u}_1^2 \mathsf{w}} \right) \Re(F_3
    F_4^*) + \left( \sqrt{3} \frac{\mathsf{u}_2^4}{\mathsf{u}_1^2
        \mathsf{v}} \right) \Re(F_L F_M^*) - \sqrt{2} \, \Re(F_1
    F_2^*)
    \Bigg) \nonumber \\
    & \quad +\frac{9 \sin{2\theta_1} \sin{2\theta_2} \sin{\phi}}{16}
    \Bigg( \left( 2 \frac{\sqrt{q_1^2 \, q_2^2}}{\mathsf{u}_1^2} \right) \Im(F_3
    F_L^*) - \left( \sqrt{3} \frac{\mathsf{u}_2^4}{\mathsf{u}_1^2
        \mathsf{v}} \right) \Im(F_3 F_M^*) - \left(
      \frac{\mathsf{u}_2^4}{\mathsf{u}_1^2
        \mathsf{w}} \right) \Im(F_4 F_L^*) \nonumber \\
    & \qquad \qquad \qquad \qquad \qquad \qquad - \left( 2 \sqrt{3}
      \sqrt{q_1^2 \, q_2^2} \frac{\mathsf{u}_2^4}{\mathsf{u}_1^2 \mathsf{v}
        \mathsf{w}} \right) \Im(F_4 F_M^*) - \left( 2 \sqrt{2} M_H
      \frac{\mathsf{u}_1}{\mathsf{v}} \right) \Im(F_1 F_M^*) \Bigg)
    \nonumber \\
    & \quad +\mathscr{M},
\end{align}
where $\mathscr{M}$ includes all the terms that are proportional to
$\eta$ and $\eta^2$ written explicitly in the appendix,
Eq.~\eqref{eq:eta-etasq}. The helicity fractions are defined as
\begin{equation}
  F_i = \frac{A_i}{\sqrt{\sum_j \modulus{A_j}^2}},
\end{equation}
and $\gammaf$is given by
\begin{equation}
  \gammaf \equiv \frac{d\Gamma}{dq_2^2} = \frac{1}{5} \; \frac{9}{2^{10}} 
  \; \frac{1}{\pi^3} \; X \; \frac{\text{Br}_{\ell\ell}^2}{M_H^2} \; 
  \frac{\Gamma_Z}{M_Z^3} \; \frac{\sum_j \modulus{A_j}^2}{\left( 
      \left( q_2^2 - M_Z^2 \right)^2 + M_Z^2 \Gamma_Z^2 \right)},
\end{equation}
where $ i, j \in \{ L,M,1,2,3,4 \}$ and we have averaged over the 5
initial polarization states of the spin-2 boson.

The uni-angular distributions are given by
\begin{align}
  \frac{1}{\gammaf} \frac{d^2 \Gamma}{dq_2^2 \; d\cos{\theta_1}} &=
  \frac{1}{2} + T_2^{(2)} \, P_2(\cos{\theta_1}) - T_1^{(2)} \,
  \cos\theta_1, \label{eq:ct1-tensor}\\
  \frac{1}{\gammaf}\frac{d^2 \Gamma}{d q_2^2 \; d\cos{\theta_2}} &=
  \frac{1}{2} + T_2^{\prime (2)} \, P_2(\cos{\theta_2}) +
  T_1^{\prime (2)} \, \cos\theta_2, \label{eq:ct2-tensor}\\
  \frac{2 \pi}{\gammaf} \frac{d^2 \Gamma}{d q_2^2 \; d\phi} &= 1 +
  U_2^{(2)} \, \cos{2\phi} + V_2^{(2)} \, \sin{2\phi} + U_1^{(2)} \,
  \cos\phi + V_1^{(2)} \, \sin\phi ,\label{eq:phi-tensor}
\end{align}
where the superscript $(2)$ is used to denote the fact that the
concerned coefficients are for spin-2 resonance, and
\begin{align}
  T_2^{(2)} &= \frac{1}{4}\bigg(-2 \modulus{F_1}^2 +\modulus{F_2}^2 +
  \left(\modulus{F_3}^2 + \modulus{F_L}^2 \right) \left(\frac{q_1^2-2
      q_2^2}{\mathsf{u}_1^2}\right) + \modulus{F_4}^2 \left(2 M_H^2
    \frac{\mathsf{u}_1^2}{\mathsf{w}^2} +
    \frac{\mathsf{u}_2^4}{\mathsf{u}_1^2 \mathsf{w}^2} \left(q_2^2-2
      q_1^2\right) \right) \nonumber \\
  & \qquad \qquad +\modulus{F_M}^2 \left(4 M_H^2
    \frac{\mathsf{u}_1^2}{\mathsf{v}^2} + 3
    \frac{\mathsf{u}_2^4}{\mathsf{u}_1^2 \mathsf{v}^2} \left(q_2^2-2
      q_1^2\right) \right) + 6 \sqrt{q_1^2 \, q_2^2}
  \frac{\mathsf{u}_2^2}{\mathsf{u}_1^2 \mathsf{v} \mathsf{w}}
  \left(\mathsf{v} \, \Re(F_3 F_4^*)+\sqrt{3} \mathsf{w} \, \Re(F_L
    F_M^*)\right) \bigg),\\
  T_2^{\prime (2)} &= \frac{1}{4} \bigg( -2 \modulus{F_1}^2 +
  \modulus{F_2}^2 + \left( \modulus{F_3}^2 + \modulus{F_L}^2 \right)
  \left(\frac{q_2^2-2 q_1^2}{\mathsf{u}_1^2}\right) + \modulus{F_4}^2
  \left(2 M_H^2 \frac{\mathsf{u}_1^2}{\mathsf{w}^2} +
    \frac{\mathsf{u}_2^4}{\mathsf{u}_1^2 \mathsf{w}^2} \left(q_1^2-2
      q_2^2\right)\right) \nonumber \\
  & \qquad \qquad + \modulus{F_M}^2 \left(4 M_H^2
    \frac{\mathsf{u}_1^2}{\mathsf{v}^2} +3
    \frac{\mathsf{u}_2^4}{\mathsf{u}_1^2 \mathsf{v}^2} \left(q_1^2-2
      q_2^2\right)\right) - 6 \sqrt{q_1^2 \, q_2^2}
  \frac{\mathsf{u}_2^2}{\mathsf{u}_1^2 \mathsf{v} \mathsf{w}}
  \left(\mathsf{v} \, \Re(F_3 F_4^*)+\sqrt{3} \mathsf{w} \, \Re(F_L
    F_M^*)\right) \bigg),\\
  U_2^{(2)} &= \frac{1}{4}\modulus{F_2}^2 - \frac{M_H^2
    \mathsf{u}_1^2}{\mathsf{v}^2} \modulus{F_M}^2,\\
  V_2^{(2)} &= M_H \frac{\mathsf{u}_1}{\mathsf{v}} \, \Im(F_2 F_M^*),\\
  T_1^{(2)} &= \frac{3 \eta}{2 \mathsf{u}_1^2 \mathsf{v} \mathsf{w}}
  \bigg( 2 M_H \mathsf{u}_1^3 \mathsf{w} \, \Re(F_2 F_M^*)+ q_1^2
  \mathsf{v} \mathsf{w} \, \Re(F_3 F_L^*) \nn\\
  & \qquad \qquad \qquad \qquad + \sqrt{q_2^2} \mathsf{u}_2^2 \left(\sqrt{3}
    \sqrt{q_1^2} \mathsf{w} \, \Re(F_3 F_M^*) + \sqrt{q_1^2} \mathsf{v} \, \Re(F_4
    F_L^*) +
    \sqrt{3} \sqrt{q_2^2} \mathsf{u}_2^2 \, \Re(F_4 F_M^*) \right) \bigg),\\
  T_1^{\prime (2)} &= \frac{3 \eta}{2 \mathsf{u}_1^2 \mathsf{v}
    \mathsf{w}} \bigg( 2 M_H \mathsf{u}_1^3 \mathsf{w} \, \Re(F_2
  F_M^*) + q_2^2 \mathsf{v} \mathsf{w} \, \Re(F_3 F_L^*) \nn\\
  & \qquad \qquad \qquad \qquad + \sqrt{q_1^2} \mathsf{u}_2^2 \left(-\sqrt{3}
    \sqrt{q_2^2} \mathsf{w} \, \Re(F_3 F_M^*) - \sqrt{q_2^2} \mathsf{v} \, \Re(F_4
    F_L^*) + \sqrt{3} \sqrt{q_1^2} \mathsf{u}_2^2
    \, \Re(F_4 F_M^*)\right) \bigg),\\
  U_1^{(2)} &= \frac{9 \pi ^2 \eta ^2}{64 \mathsf{u}_1^2 \mathsf{v}^2
    \mathsf{w}^2} \bigg(\sqrt{2} \mathsf{u}_1^2 \mathsf{v}^2
  \mathsf{w}^2 \, \Re(F_1 F_2^*)-\mathsf{u}_2^4 \mathsf{v}^2
  \mathsf{w} \, \Re(F_3 F_4^*)+\modulus{F_3}^2 \sqrt{q_1^2 \, q_2^2} \mathsf{v}^2
  \mathsf{w}^2-\modulus{F_4}^2 \sqrt{q_1^2 \, q_2^2} \mathsf{u}_2^4 \mathsf{v}^2
  \nn\\
  & \qquad \qquad \qquad \qquad +\sqrt{3} \mathsf{u}_2^4 \mathsf{v}
  \mathsf{w}^2 \, \Re(F_L F_M^*)-\modulus{F_L}^2 \sqrt{q_1^2 \, q_2^2} \mathsf{v}^2
  \mathsf{w}^2+3 \modulus{F_M}^2 \sqrt{q_1^2 \, q_2^2} \mathsf{u}_2^4
  \mathsf{w}^2\bigg),\\
  V_1^{(2)} &=\frac{9 \pi ^2 \eta ^2}{64 \mathsf{u}_1^2 \mathsf{v}
    \mathsf{w}} \bigg(2 \sqrt{2} M_H \mathsf{u}_1^3 \mathsf{w} \,
  \Im(F_1
  F_M^*)+2 \sqrt{q_1^2 \, q_2^2} \mathsf{v} \mathsf{w} \, \Im(F_3 F_L^*)\nn\\
  & \qquad \qquad \qquad \qquad - \sqrt{3} \mathsf{u}_2^4 \mathsf{w}
  \, \Im(F_3 F_M^*) - \mathsf{u}_2^4 \mathsf{v} \, \Im(F_4 F_L^*) - 2
  \sqrt{3} \mathsf{u}_2^4 \sqrt{q_1^2 \, q_2^2} \, \Im(F_4 F_M^*) \bigg).
\end{align}
\end{widetext}
These coefficients can again be extracted from asymmetries similar to
those defined in Eqs.~\eqref{eq:T10asym}, \eqref{eq:T20asym},
\eqref{eq:U10asym}, \eqref{eq:U20asym}, \eqref{eq:V10asym} and
\eqref{eq:V20asym} for the spin-0 case.
We find that the angular distributions corresponding to $P_2
(\cos\theta_1)$ and $P_2 (\cos\theta_2)$ are different in the Spin-2
case in contrast to the Spin-0 case. This feature can enable us to
distinguish between the two spins, unless the difference happens to be
zero for certain choice of parameters, even in the Spin-2
case. Considering only the $\eta$ independent terms in
Eqs.~\eqref{eq:ct1-tensor} and \eqref{eq:ct2-tensor}, the difference
$\Delta$ between the coefficients of $P_2(\cos{\theta_1})$ and
$P_2(\cos{\theta_2})$ in $\displaystyle \frac{1}{\gammaf} \frac{d^2
  \Gamma}{dq_2^2 \; d\cos{\theta_1}}$ and $\displaystyle
\frac{1}{\gammaf} \frac{d^2 \Gamma}{dq_2^2 \; d\cos{\theta_2}}$
respectively, is
\begin{align}
  \Delta &= \frac{3\mathsf{u}_2^2}{4 \mathsf{u}_1^2 \mathsf{v}^2
    \mathsf{w}^2} \bigg( \mathsf{v}^2 \mathsf{w}^2 \left(
    \modulus{F_3}^2 + \modulus{F_L}^2 \right) \nonumber \\
  & \qquad \qquad \qquad \quad - \mathsf{u}_2^4 \left( \mathsf{v}^2
    \modulus{F_4}^2 + 3 \, \mathsf{w}^2 \modulus{F_M}^2 \right) \bigg)
  \nonumber \\
  &+\frac{3\sqrt{q_1^2 \, q_2^2} \mathsf{u}_2^2}{\mathsf{u}_1^2 \mathsf{v}
    \mathsf{w}} \bigg( \mathsf{v} \, \Re{(F_3 F_4^*)} + \sqrt{3} \,
  \mathsf{w} \, \Re{(F_L F_M^*)}\bigg). \label{eq:Delta}
\end{align}
If we find that $\Delta = 0$ for all $\sqrt{q_2^2}$, then the tensor
case would have similar characteristics in the uni-angular
distributions as discussed in the scalar case.  However, this can only
happen if helicity amplitudes (or equivalently the corresponding
coefficients $A$, $B$, $C$, $D$, $E$ and $F$) have the explicit
momentum dependence so as to absorb $\sqrt{q_2^2}$ completely in
$\Delta$. The reader can examine the expression for $\Delta$ to
conclude that this is impossible and the only way $\Delta$ can be
equated to zero for all $\sqrt{q_2^2}$, is when
\begin{equation}
  F_3 = F_4 = F_L = F_M =0.
\end{equation}
In such a special case all the form-factors in vertex
$V^{\mu\nu;\alpha\beta}_{HZZ}$ vanish, except $C$ and $D$.  This
special case explicitly implies that the parity of the Spin-2 boson is
even. We will refer to this case as the special $J^P=2^+$ case, since
the uni-angular distribution mimics the $J^P=0^+$ case. Working under
this special case
\begin{align}
  \frac{1}{\gammaf} \frac{d^2 \Gamma}{dq_2^2 \; d\cos{\theta_1}} &=
  \frac{1}{2} + T_2^{(2)} P_2(\cos{\theta_1}), \label{eq:ct1-tensor-2++}\\
  \frac{1}{\gammaf}\frac{d^2 \Gamma}{dq_2^2 \; d\cos{\theta_2}} &=
  \frac{1}{2} + T_2^{(2)} P_2(\cos{\theta_2}), \label{eq:ct2-tensor-2++}\\
  \frac{2 \pi}{\gammaf} \frac{d^2 \Gamma}{dq_2^2 \; d\phi} &= 1 +
  U_2^{(2)} \cos{2\phi} + U_1^{(2)} \cos\phi,
\label{eq:phi-tensor-2++}
\end{align}
where the $T_2^{(2)}$, $U_2^{(2)}$ and $U_1^{(2)}$ are now given by
\begin{align}
  T_2^{(2)} &= \frac{1}{4} \; \left( \modulus{F_2}^2 - 2 \modulus{F_1}^2 \right),\\
  U_2^{(2)} &= \frac{1}{4} \modulus{F_2}^2, \\
  U_1^{(2)} &= \frac{9\pi^2}{32\sqrt{2}}\, \eta^2 \, \Re(F_1 F_2^*)
\end{align}
Now using the identity $\modulus{F_1}^2 + \modulus{F_2}^2 = 1$, we get
\begin{equation}\label{eq:T1T2-tensor}
  U_2^{(2)} = \frac{1}{6} \left( 1 + 2 T_2^{(2)} \right).
\end{equation}
Note the similarity between Eqs.~\eqref{eq:T2-scalar} and
\eqref{eq:T1T2-tensor}. The conclusions that $J^P=2^\pm$ when
$\Delta\neq 0$ can also be drawn if $\Delta$ integrated over $q_1^2$
and $q_2^2$ is found to be non zero. However, it clear from
Eq.~\eqref{eq:Delta} that the domain of integration for $q_1^2$ and
$q_2^2$ cannot be symmetric.

\begin{figure}[hbtp]
\centering
\includegraphics{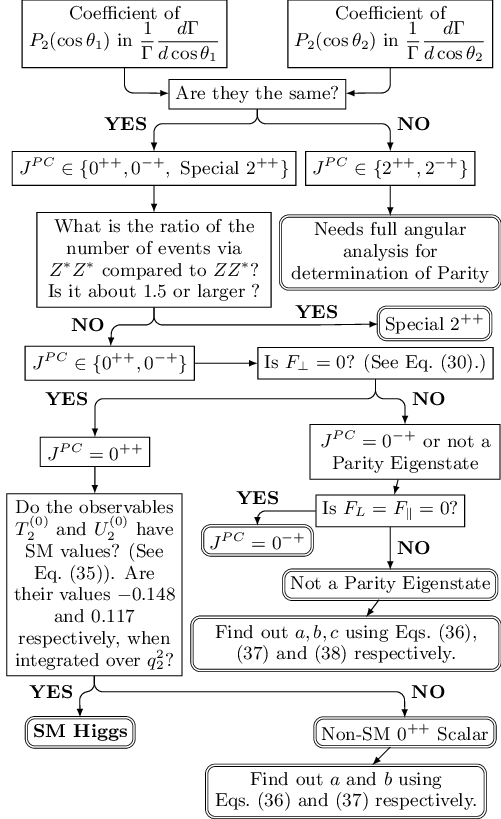}
\caption{Flow chart for determination of spin and parity of the new
  boson. See text for details.}
\label{fig:flowchart}
\end{figure}

\subsection{Comparison Between Spin-0 and Spin-2}\label{subsec:comparison}
Having discussed both the scalar and tensor case, we summarize the
procedure to distinguish the spin and parity states of the new boson
in a flowchart in Fig.~\ref{fig:flowchart}. The procedure entailed,
ensures that we convincingly determine the spin and parity of the
boson. The first step should be to compare the uni-angular
distributions in $\cos{\theta_1}$ and $\cos{\theta_2}$. If the
distribution is found to be different the boson cannot be the SM Higgs
and indeed must have Spin-2. However, if the distributions are found
to be identical the resonance can have Spin-0 or be a very special
case of Spin-2 arising only from $C$ and $D$ terms in the vertex in
Eq.~\eqref{eq:VHZZ2}.  The similarity between
Eqs.~\eqref{eq:T2-scalar} and \eqref{eq:T1T2-tensor} makes it
impossible to distinguish these two cases by looking at angular
distributions alone.

The special $J^P=2^+$ case can nevertheless still be identified by
examining the surviving helicity amplitudes $A_1$ and $A_2$. The
helicity amplitudes given in Eqs.~\eqref{eq:A1} and \eqref{eq:A2}
reduce in this special case to,
\begin{align}
  A_1 &= -\frac{16 \sqrt{2}}{3 \sqrt{3}} \; X^2\; \bigg(q_1.q_2 \, C +
  M_H^2 X^2 \, D \bigg),\label{eq:A1-new}\\
  A_2 &= \frac{32}{3 \sqrt{3}}\; \sqrt{q_1^2 \, q_2^2} \, X^2 \,
  C. \label{eq:A2-new}
\end{align}
These may be compared with Eqs.~\eqref{eq:AL} and \eqref{eq:AA} to
notice that they have identical form, except for an additional $X^2$
dependence in $A_1$ and $A_2$ expressions above. The additional $X^2$
dependence increases the contribution from both off-shell $Z$'s
(called $Z^*Z^*$) significantly in comparison to the dominant one
on-shell and one off-shell $Z$ (called $Z Z^*$) contribution expected
in SM. In the SM one would expect the ratio of the number of events in
$Z^*Z^*$ to $ZZ^*$ channel to be about $0.2$. However, in the special
$J^P=2^+$ case we would expect this ratio to be about $1.5$.  Thef
reader is cautioned not to confuse this explicit $X^2$ dependence with
any assumption on the momentum dependence of the
form-factors. Throughout the analysis we have assumed the most general
form-factors $a$, $b$, $c$, $A$, $B$, $C$, $D$, $E$ and $F$,
nevertheless $A_1$ and $A_2$ turn out to have additional $X^2$
dependence in comparison to $A_L$ and $A_\|$ respectively. This
explicit $X^2$ dependence arises due to contributions only from higher
dimensional operators in the special $J^P=2^+$ case.

Having excluded the Spin-2 possibility, the resonance would be a
parity-odd state ($0^{-+}$) if $F_L = F_{\parallel} = 0$ and a
parity-even state ($0^{++}$) if $F_{\perp}=0$. If the resonance is
found to be in $0^{++}$ state, we need to check whether $T_2^{(0)}$
and $U_2^{(0)}$ terms are as predicted in SM. The values of
$T_2^{(0)}$ and $U_2^{(0)}$ as a function of $\sqrt{q_2^2}$ are
plotted in Fig.~\ref{fig:SM-Higgs}. The $q_2^2$ integrated values for
the observables $T_2^{(0)}$ and $U_2^{(0)}$ are uniquely predicted in
SM at tree level to be $-0.148$ and $0.117$ respectively.  These tests
would ascertain whether the $0^{++}$ state is the SM Higgs or some
non-SM boson. If it turns out to be a non-SM boson, we can also
measure the coefficients $a,b,c$ by using Eqs.~\eqref{eq:a},
\eqref{eq:b} and \eqref{eq:c}.

Finally we emphasize that our approach is unique in using helicity
amplitudes in the transversity basis so that the amplitudes are
classified by parity. We also use orthogonality of Legendre
polynomials in $\cos\theta_1$ and $\cos\theta_2$ as well as a Fourier
series in $\phi$ to unambiguously determine the spin and parity of the
new resonance.  Another significant achievement is the use of the most
general $HZZ$ vertex factors for both Spin-0 and Spin-2 cases allowing
us to determine the nature of $H$ be it in any extension of the SM. We
wish to stress that we consider neither any specific mode of
production of the new resonance (like gluon-gluon fusion or vector
boson fusion), nor any specific model for its couplings. The
production channel for the new resonance has no role in our
analysis. We consider its decay only to four leptons via two $Z$
bosons. Most discussions in current literature deal either with
specific production channels or with specific models of new physics
which restrict the couplings to specific cases both for Spin-0 and
Spin-2.  Refs.~\cite{Ellis:2012wg, Ellis:2012jv, Ellis:2012xd,
  Ellis:2012mj, Geng:2012hy} deal with graviton-like Spin-2 particles,
while Ref.~\cite{Frank:2012wh} deals with Spin-2 states that are
singlet or triplet under $SU(2)$. Ref.~\cite{Ellis:2012wg} considers
polar angle distribution of $\gamma\gamma$ and angular correlations
between the charged leptons coming from $WW^*$ decays to differentiate
the Spin-0 and Spin-2 possibilities.  While Ref.~\cite{Ellis:2012xd}
looks at `Higgs'-strahlung process to distinguish the various spin and
parity possibilities, Ref.~\cite{Ellis:2012jv} compares branching
ratios of the new boson decaying to $\gamma\gamma$, $WW^*$ and $ZZ^*$
channels as a method to measure the spin and parity of the new
boson. In Ref.~\cite{Geng:2012hy} the authors propose a new observable
that can distinguish SM Higgs from a Spin-2 possibility. They consider
the three-body decay of the new resonance to a SM vector boson and a
fermion-antifermion pair.  Ref.~\cite{Ellis:2012mj} shows that the
current data disfavors a particular type of graviton-like Spin-2
particle that appears in scenarios with a warped extra dimension of
the AdS type. Refs.~\cite{Frank:2012wh,Djouadi:2013-21-Jan} deal with
Spin-0 or Spin-2 particles produced via vector boson fusion process
alone. Our discussion subsumes all of the above special cases.
Moreover, unlike other discussions in the literature we provide
clearly laid out steps to measure the couplings, spin and parity of
the new resonance $H$ without any ambiguity. We want to reiterate that
it is important to measure not only the spin and parity of the new
resonance but also its couplings before any conclusive statements can
be made that it is the SM Higgs.

\subsection{Numerical study of the uni-angular
  distributions}\label{subsec:numerical}

In this sub-section we study the possibility of using the uni-angular
distributions, given in previous sub-section to differentiate the
different possible spin $CP$ states. For simplicity throughout this
sub-section we will neglect the $q^2$ dependence of $a,b$ and $c$.
The signal and background events were generated using the
MadEvent5~\cite{Alwall:2011uj} event generator interfaced with PYTHIA
6.4~\cite{Sjostrand:2003wg} and PGS 4~\cite{pgs}. The vertex of
Eq.~(\ref{eq:VHZZ0}) was implemented into the UFO format of Madgraph5
using Feynrules 1.6.18~\cite{Alloul:2013bka}. Unlike the earlier
sub-sections we also include the $2e^+ 2e^-$ and $2\mu^+ 2\mu^-$ final
states because the identification of $Z_1$ being the mother particle
of the pair of same flavor opposite sign leptons with an invariant
mass closest to the $M_Z$ breaks the exchange symmetry of these final
states in most regions of phase space. As the analysis of this paper
has to do purely with the shape of the partial widths in the
$Z^{(*)}Z^{(*)}$ channel, the production mechanism is not crucial to
understanding the spin and $CP$ properties of the resonance at
$125~\gev$. However to be concrete, these samples were generated for
$pp$ collisions at $\sqrt{s} = 8~\tev$ using the CTEQ6L1 parton
distribution functions (PDFs)~\cite{Pumplin:2002vw}. We choose to
follow the ATLAS cut based analysis of Ref.~\cite{atlash2zz} instead
of the CMS analysis~\cite{Chatrchyan:2012jja} because the CMS analysis
has used a more sophisticated multi-variate analysis (MVA)
technique. We set the Higgs boson mass $m_H = 125~\gev$, which is
close to what has been measured in Ref.~\cite{atlash2zz}. The
branching ratios and decay widths are set appropriately using the
values from the Higgs working group webpage~\cite{hwgpage}.

Following the analysis of Ref.~\cite{atlash2zz} we impose the
following lepton selection cuts and triggers. In particular, the
single lepton trigger thresholds are $p_T^l > 24 (25)~\gev$ for a
muon(electron). The di-muon trigger thresholds used are $p_T >
13~\gev$ for the symmetric case and $p_T^1 > 18~\gev$ and $p_T^2 >
8~\gev$ for the asymmetric case. For di-electrons the thresholds are
$p_T > 12~\gev$. The lepton identification cuts require that each
electron(muon) must have $E_T > 7~\gev$ ($p_T > 6~\gev$) with $|\eta|
< 2.4 (2.7)$. Sorting leptons in decreasing order of $p_T$, we also
impose the selection criteria $p_T^{\ell_1} > 20~\gev$, $p_T^{\ell 2}
> 15~\gev$ and $p_T^{\ell_3} > 10~\gev$. For same flavor leptons we
also require that $\Delta R > 0.1$ while for opposite flavor $\Delta R
> 0.2$. Furthermore we also impose the invariant mass cuts on the
$m_{Z_1}$, $m_{Z_2}$ and $m_{4\ell}$ described in
Table~\ref{cut_table} to reduce the Standard Model
background. $m_{Z_1}$ is the invariant mass of the pair of opposite
sign same flavor leptons closest to $m_Z$ while $m_{Z_2}$ is the other
combination. The number of signal events in our simulation is in good
agreement with the SM predicted value quoted in Ref.~\cite{atlash2zz},
while the background rate is slightly lower than total background rate
because we have not included the sub-dominant processes like Z+jets
and $t \bar t$.

\begin{table}[hbtp]
\centering
\begin{tabular}{|c|c|c|}
\hline
Cuts & $m_H = 125~\gev$& SM $ZZ^*$ \\
\hline
Selection & 22 & 1542 \\
$50\mbox{ GeV} < m_{Z_1} < 106$~GeV & 20 &  1432 \\
$12\mbox{ GeV} < m_{Z_2} < 115$~GeV & 19 &  1294 \\
$115 \mbox{ GeV} < m_{4\ell} < 130$~GeV & 19 &  14 \\
\hline
\end{tabular}
\caption{Effect of the sequential cuts on the simulated Signal and the
  dominant continuum $ZZ$ background, where the $k$-factors are $1.3$ for
  signal and $2.2$ for background using MCFM 6.6~\cite{mcfm} for 20.7
  fb$^{-1}$.}
\label{cut_table}
\end{table}

In order to quantify the effect of using the uni-angular distributions
to extract the nature of the $125$~GeV resonance we construct the test
statistic $q$ based on the ratio of the likelihoods
\begin{equation}
  q = \ln \frac{\mathcal{L}_{0^+}}{\mathcal{L}_{0^-}}~,
  \label{eq:qstat}
\end{equation}
where the $\mathcal{L}$ is the unbinned likelihood function
\begin{equation}
\mathcal{L} = \sum_{\mu_s} \left(\prod_{i}^{N_{\rm obs}} \frac{\mu_s
  P_s (\mathbf{x}_i) + \mu_b P_b (\mathbf{x}_i)}{\mu_s + \mu_b}
\right)_{\rm ave}.
\label{eq:likelihood}
\end{equation}
As our acceptances are in good agreement with the ATLAS predictions
for the rest of our analysis we will assume a background rate $\mu_b
=16$ events for luminosity $L = 20.7\mbox{ fb}^{-1}$ due to the
continuum $ZZ$ background. However as the total observed number of
events are slightly above the expected rate we need to marginalize
over the expected signal rate. In particular we assume a bayesian
prior flat distribution for $\mu_s \in [0.5,2.0] \times \mu_s^{\rm
  SM}(= 18 \mbox{ at a luminosity of } \, 20.7 {\rm fb}^{-1})$. For a
particular value of $\mu_s$ we generate ensembles of $N_{\rm obs}$
events to find the average of the product within the brackets in
Eq.~\eqref{eq:likelihood}. The probability density function (PDF) for
signal is the product of the distributions
\begin{widetext}
\begin{align}
\frac{1}{\Gamma}\frac{d\Gamma}{d\cos\theta_1} &= \frac{1}{2} -
\mathcal{T}_1^{(0)}(a,B,C) \, \cos\theta_1 +
\mathcal{T}_2^{(0)}(a,B,C) \, P_2(\cos\theta_1),\label{eq:Ct1} \\
\frac{1}{\Gamma}\frac{d\Gamma}{d\cos\theta_2} &= \frac{1}{2} +
\mathcal{T}_1^{(0)}(a,B,C) \, \cos\theta_2 +
\mathcal{T}_2^{(0)}(a,B,C) \, P_2(\cos\theta_2),\label{eq:Ct2}\\
\frac{1}{\Gamma}\frac{d\Gamma}{d\phi} &= \frac{1}{2\pi} +
\mathcal{U}_1^{(0)}(a,B,C) \, \cos\phi + \mathcal{U}_2^{(0)}(a,B,C)
\,\cos2\phi,\label{eq:Phi}
\end{align}
where $B=b \times (100~\gev)^2$, $C=c\times (100~\gev)^2$ and
\begin{align}
  \Gamma \equiv \Gamma(a,B,C) &\simeq 2.24 \times 10^{-8} \,x_H^{14}
  \big( a^2 +0.19 \,a \,B + 2.22\times10^{-2} \,B^2 \,x_H^2 + 2.14\times10^{-2} \,C^2
  \,x_H^6 \big), \label{eq:decaywidth} \\
  \mathcal{T}_1^{(0)}(a,B,C) &\simeq \frac{2.14\times10^{-2} \,a \,C \,x_H^3}{a^2
    +0.19 \,a \,B + 2.22\times10^{-2} \,B^2 \,x_H^2 + 2.14\times10^{-2} \,C^2
    \,x_H^6},\label{eq:t10} \\
  \mathcal{T}_2^{(0)}(a,B,C) &\simeq \frac{-0.15\,a^2 -
    9.65\times10^{-2}\,a\,B\,x_H^3 + 5.35\times10^{-3}\,C^2\,x_H^9}{x_H^3 \left(a^2 + 0.19
    \,a \,B + 2.22\times10^{-2} \,B^2 \,x_H^2 + 2.14\times10^{-2} \,C^2 \,x_H^6
    \right)},\label{eq:t20}\\
  \mathcal{U}_1^{(0)}(a,B,C) &\simeq \frac{-3.44\times10^{-3}\,a^2 -
    5.50\times10^{-4}\,a\,B\,x_H^2}{a^2 + 0.19 \,a \,B + 2.22\times10^{-2} \,B^2 \,x_H^2 +
    2.14\times10^{-2} \,C^2 \,x_H^6},\label{eq:u10}\\
  \mathcal{U}_2^{(0)}(a,B,C) &\simeq \frac{1.88\times10^{-2}\,a^2 \, x_H -
    8.51\times10^{-4}\,C^2\,x_H^6}{a^2 +0.19 \,a \,B + 2.22\times10^{-2} \,B^2 \,x_H^2 +
    2.14\times10^{-2} \,C^2 \,x_H^6},\label{eq:u20}
\end{align}
\end{widetext}
while for the background $P_b = 1/(8\pi)$. In the above approximations
for we have neglected the $q^2$ dependences of $a,b$ and $c$ and
integrated Eq~(\ref{eq:gamma}), Eq.~(\ref{eq:scalar-costheta1}),
Eq.~(\ref{eq:scalar-costheta2}) and Eq.~(\ref{eq:scalar-phi}) over
$q_2^2$. Furthermore we have performed a power law fit in term of $x_H
= m_H/(120~\gev)$ for each of the coefficients. As $b$ and $c$ have
dimensions of mass squared, in the above approximations for the
different coefficients we have used the dimensionless coefficients $B$
and $C$ instead. By definition, the $0^+$ hypothesis corresponds to
$(a,B,C) = (1,0,0)$ and the $0^-$ hypothesis corresponds to $(a,B,C) =
(0,0,1)$. When $a = 0$ the magnitude of $C$ is not crucial as we
normalize the $0^+$ and $0^-$ cross-sections so as to produce the same
number of signal events.

\begin{figure}
\centering
\includegraphics[width=0.45\textwidth]{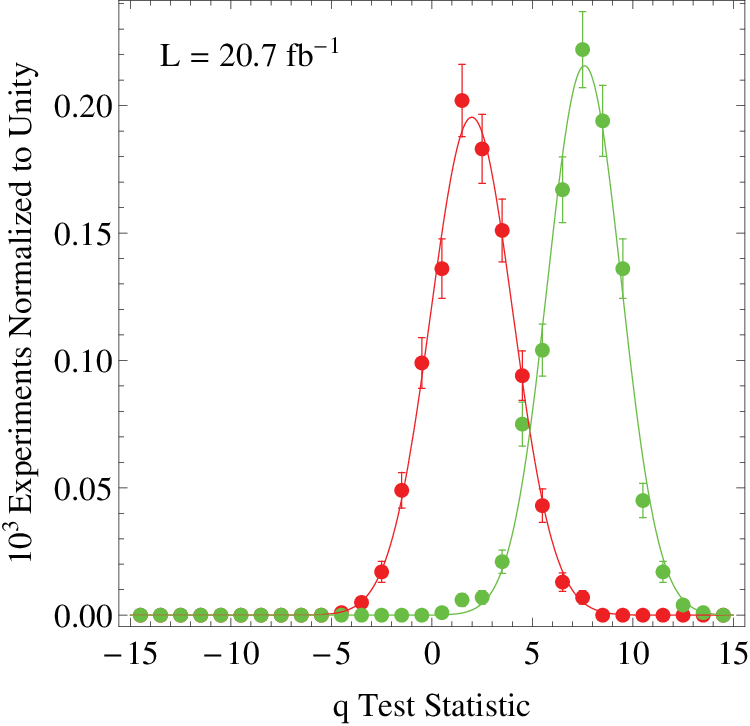}
\caption{Comparison of the q test-statistic using the uni-angular
  distribution approach in the 4$\ell$ channel for the $0^+$ events in
  red (gray) vs. $0^-$ events in green (light gray).}
\label{fig:qvsm}
\end{figure}

\begin{figure}
\centering
\includegraphics[width=0.45\textwidth]{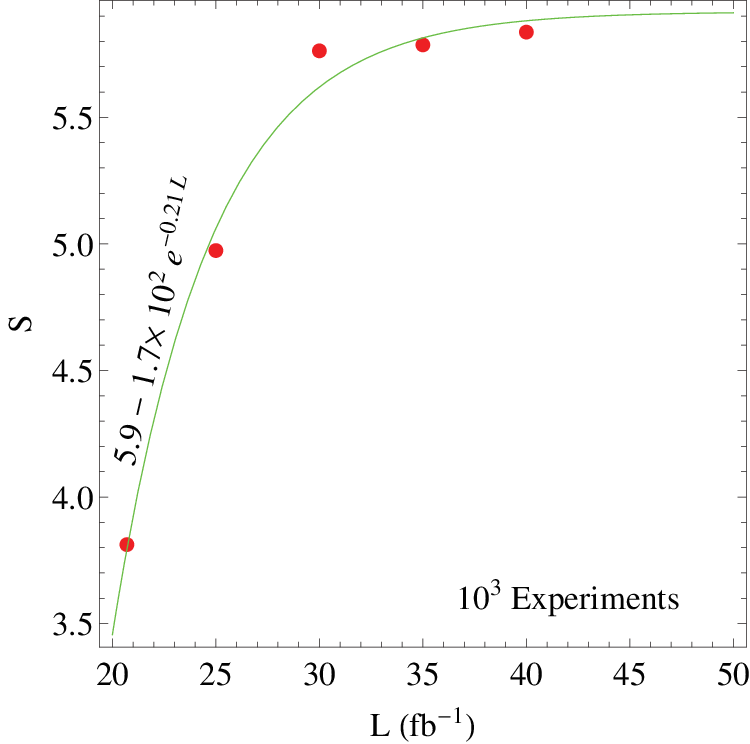}
\caption{Separation power for $q$-test statistic using the uni-angular
  distributions as a function of Luminosity. The red (dark grey)
  points are the simulated separation power and the green (light grey)
  curve is the fit to the data}
\label{fig:sep_pow}
\end{figure}

To quantify power of the uni-angular distributions in hypothesis
testing, we present the q test-statisic for the $0^+$ and $0^-$
hypotheses in Fig.~\ref{fig:qvsm}. In particular, we have applied the
q-statistic in Eq.~(\ref{eq:qstat}) to samples of Monte Carlo events
that have passed the above cuts in Tab.~\ref{cut_table}, where we
assumed the above bayesian prior for the mean signal rate. The red
(dark grey) curve corresponds to $0^-$ events while the green (light
grey) curve corresponds to $0^+$ events. The solid curves correspond
to a gaussian fit to these distributions and using them we define the
separation power as
\begin{equation}
S = \frac{2 A}{\sigma},
\end{equation} 
where $A$ is the area under the curve calculated from the point on the
$q$-axis which satisfies the condition that the area under the right
tail of the left distribution is equal to the left tail of the right
distribution and $\sigma$ is the maximum of the two standard
deviations.

The separation power using the q test statistic works well for low
luminosity, but this approach loses sensitivity at larger luminosity. To
illustrate this point we present Fig.~\ref{fig:sep_pow} at a function
of luminosity. The red (dark grey) points correspond calculated
separation power for a particular luminosity while the green (light
grey) curve is a fit to the data. The lowest data point corresponds to
a luminosity of $20.7\mbox{ fb}^{-1}$ with an observation of 43 events
while for at higher luminosities we have assumed that the number of
observed events agrees with the expected rates. Furthermore this
extrapolation assumes the same cuts and efficiencies for higher
luminosities. For luminosities greater that $40\mbox{ fb}^{-1}$, a
$\chi^2$ fit of the uni-angular distributions would probably provide a
stronger hypothesis test.

\begin{figure}
\centering
\includegraphics[width=0.40\textwidth]{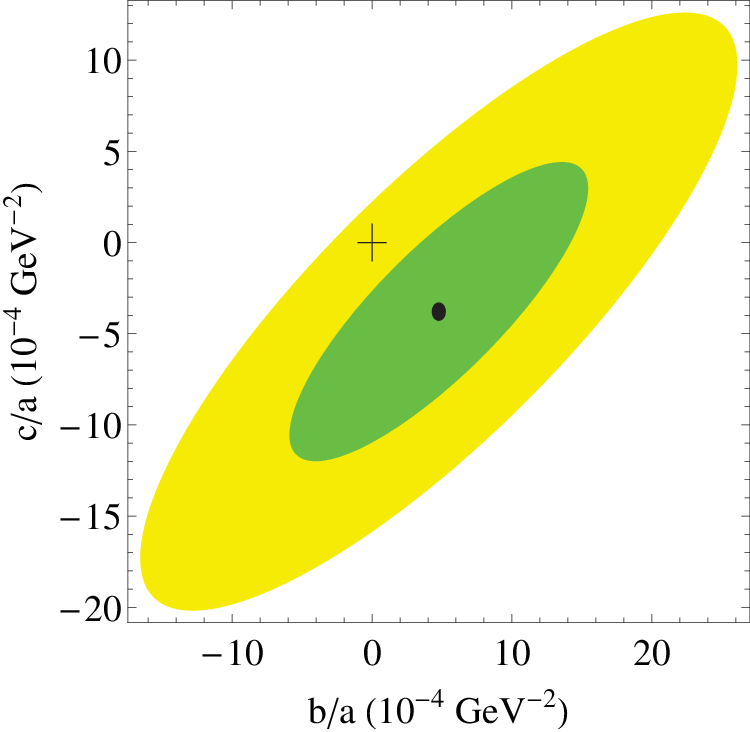} 
\caption{$c/a$ vs $b/a$ $1\sigma$ (green) and $2\sigma$ (yellow)
  contours assuming the Standard Model value of the partial decay
  width to 4$\ell$.  The central values $(b/a, c/a) =
  (4.77 \pm 21.23,-3.79 \pm 16.4) \times 10^{-4}~{\rm GeV}^{-2}$ is
  shown by the block dot. The cross-hair corresponds to $b = c = 0$.}
\label{fig:cvsbplot}
\end{figure}

It would seem that the values of all the form factors $a$, $b$ and $c$
can be extracted using the there uni-angular
distributions~Eq.~\eqref{eq:Ct1}\--\eqref{eq:Phi} along with
Eq.~(\ref{eq:decaywidth})\--(\ref{eq:u20}). However, the difference
between the uni-angular distributions in Eq.~(\ref{eq:Ct1}) and
Eq.~(\ref{eq:Ct2}) is small because it is proportional to $\eta$.
Given the small sample of 43 events this would essentially imply that
only two parameters can be obtained. Our numerical work confirms this
fact. Since $P_0(\cos \theta_{1,2})=1$, $P_1(\cos \theta_{1,2})$,
$P_2(\cos \theta_{1,2})$, $\cos \phi$ and $\cos 2\phi$ are orthogonal
functions the coefficients of each of the terms can be extracted
individually. As discussed in Sec.~\ref{subsec:scalar} this would
result in four observables. We emphasize that as the data sample
increases the additional information can be used to measure relative
phases between $a$, $b$ and $c$. For 43 events, as expected from the
discussions in Sec.~\ref{subsec:scalar} based on the small value of
$\eta$ in SM, we find we could only extract stable values of $b/a$ and
$c/a$ by maximizing the likelihood function $ \mathcal{L}_{0^+}$. One
can also estimate the errors in $b/a$ and $c/a$ from the inverse of
the covariance matrix $V_{ij}= cov[\theta_i , \theta_j]$ defined as
\begin{eqnarray}
\hat{V}^{-1}=-\bigg(\frac{\partial^2 \ln\mathcal{L}}{\partial \theta_i
  \partial \theta_j}\bigg)_{\hat{\theta}} \label{eq:invcov-matrix}
\end{eqnarray}
where $\theta_i,\theta_j = b/a,c/a$. Here $\hat{\theta}$ denotes those
values of the parameters that maximizes the likelihood function. In
Fig.~\ref{fig:cvsbplot} we present the extract values of $b/a$ and
$c/a$ for a sample of 43 events. 
Using these values of $b/a$ and $c/a$, the value of $a$ can
also be found by fitting the decay width in Eq.~(\ref{eq:decaywidth})
to the Standard Model partial width.  Using this approach, the values of $a$, $b$ and $c$
with their respective errors are
\begin{align}
  a &= 2.11 \pm 3.55, \\
  b &= (10.09 \pm 47.99) \times 10^{-4}~{\rm \gev}^{-2}, \\
  c &= -(8.01 \pm 37.20) \times 10^{-4}~{\rm \gev}^{-2}.
\end{align}

\section{CONCLUSION}\label{sec:conclusion}

We conclude that by looking at the three uni-angular distributions and
examining the numbers of $Z^*Z^*$ to $ZZ^*$ events one can
unambiguously confirm whether the new boson is indeed the Higgs with
$J^{PC}=0^{++}$ and with couplings to $Z$ bosons exactly as predicted
in the Standard Model.  We show that the terms in the angular
distribution corresponding to $P_2(\cos{\theta_1})$ and
$P_2(\cos{\theta_2})$ play a critical role in distinguishing the $J=2$
and $J=0$ states.  The distributions are identical for Spin-0 case,
but must be different for Spin-2 state except in a special $J^P=2^+$
case where $F_3 = F_4 = F_L = F_M =0$.  The ratio of the number of
$Z^*Z^*$ events to the number of $ZZ^*$ events provides a unique
identification for this special $J^P=2^+$ case. In this special case
the number of $Z^*Z^*$ events dominates significantly over the number
of $ZZ^*$ events. The Spin-2 resonance can thus be unambiguously
confirmed or ruled out. With Spin-2 possibility ruled out, Spin-0 can
be studied in detail.

The resonance would then be a parity-odd state ($0^{-+}$) if $F_L =
F_{\parallel} = 0$ and a parity-even state ($0^{++}$) if
$F_{\perp}=0$. If the resonance is found to be in $0^{++}$ state, we
need to check whether $T_2^{(0)}$ and $U_2^{(0)}$ terms are as
predicted in SM. The $q_2^2$ integrated values for the observables
$T_2^{(0)}$ and $U_2^{(0)}$ are uniquely predicted in SM at tree level
to be $-0.148$ and $0.117$ respectively.  These tests would ascertain
whether the $0^{++}$ state is the SM Higgs or some non-SM boson.  If
it turns out to be a non-SM boson, we can also measure the
coefficients $a,b,c$ by using Eqs.~\eqref{eq:a}, \eqref{eq:b} and
\eqref{eq:c}. If the boson is a mixed parity state, the relative phase
between the parity-even and parity-odd amplitudes can also be measured
by studying the $\sin{2\phi}$ term in the uni-angular distribution.
We present a step by step methodology in Fig.~\ref{fig:flowchart} for
a quick and sure-footed determination of spin and parity of the newly
discovered boson.  Our approach of using Legendre polynomials and the
choice of helicity amplitudes classified by parity enable us to
construct angular asymmetries that
unambiguously determine if the new resonance is indeed the Standard
Model Higgs.

Numerically we have have simulated the dominant continuum $ZZ$
background and Standard Model signal shown that our acceptances are in
good agreement with the ATLAS predictions. Using the uni-angular
distributions derived in this paper we compute the q-statistic $q
=\ln\left( \mathcal{L}_{0^+}/\mathcal{L}_{0^-} \right)$.  We observe
the separation power of this approach is most powerful at low
luminosity assuming that the cuts and the acceptances remain the same
at each luminosity.
For easy experimental adaption we have included power law
parametrization of the various angular coefficients in terms of the
fundamental Higgs vertex parameters. 
We also obtain fits for $b/a$ and $c/a$ for a 43-event sample,
demonstrating that both $b$ and $c$ can be constrained by a
rather small sample of data.  \vspace{1cm}
\noindent
\acknowledgments
RS is grateful to Institute of Physics, Academia Sinica, for
hospitality where part of the work was done. We thank Sridhara Dasu
for discussions. AM is supported by the U.S. Department of Energy
under Contract No. DE-FG02-96ER40969.

\appendix
\section{Other Terms in the Angular Distributions}\label{sec:extra-terms} 
In the main text, we have not included the $\eta$ and $\eta^2$
dependent term in the angular distributions for the case of Spin-2
boson. However, for the sake of completeness, the $\eta$ and $\eta^2$
dependent term $\mathscr{M}$ in the angular distributions are given
below.

\begin{widetext}
\begin{align}
  \mathscr{M} &= \eta \Bigg( -3 M_H \Re(F_2 F_M^*)
  \frac{\mathsf{u}_1}{\mathsf{v}} \left( \cos{\theta_1}
    (P_2(\cos{\theta_2}) + 2) - \cos{\theta_2} (P_2(\cos{\theta_1})+2)
  \right) \nonumber \\
  & \qquad -\frac{3}{\mathsf{u}_1^2} \Re(F_3 F_L^*) \left( q_1^2
    \cos{\theta_1} (1 - P_2(\cos{\theta_2})) - q_2^2 \cos{\theta_2} (
    1
    - P_2(\cos{\theta_1}) ) \right) \nonumber \\
  & \qquad -3 \sqrt{3} \sqrt{q_1^2 \, q_2^2} \Re(F_3 F_M^*)
  \frac{\mathsf{u}_2^2}{\mathsf{u}_1^2 \mathsf{v}} \left(
    \cos{\theta_1} (1-P_2(\cos{\theta_2})) + \cos{\theta_2}
    (1-P_2(\cos{\theta_1}))
  \right) \nonumber \\
  & \qquad -3 \sqrt{q_1^2 \, q_2^2} \Re(F_4 F_L^*)
  \frac{\mathsf{u}_2^2}{\mathsf{u}_1^2 \mathsf{w}} \left(
    \cos{\theta_1} (1-P_2(\cos{\theta_2})) + \cos{\theta_2}
    (1-P_2(\cos{\theta_1}))
  \right) \nonumber \\
  & \qquad +12 \sqrt{3} \mathsf{u}_2^4 \Re(F_4 F_M^*) \frac{1}{4
    \mathsf{u}_1^2 \mathsf{v}^3 \mathsf{w}^3} \bigg( - q_2^2
  \mathsf{v}^2 \mathsf{w}^2 \cos{\theta_1} (1-P_2(\cos{\theta_2}))
  \nonumber \\
  & \qquad \qquad \qquad \qquad \qquad \qquad \qquad \qquad + q_1^2
  \cos{\theta_2} \left( \mathsf{v}^2 \mathsf{w}^2 -
    P_2(\cos{\theta_1}) \left(8 M_H^4 \mathsf{u}_1^4+10 M_H^2
      \mathsf{u}_1^2
      \mathsf{u}_2^4+3 \mathsf{u}_2^8\right) \right) \bigg) \nonumber \\
  & \qquad +(\sin{\theta_1} \sin{\theta_2} \sin{\phi}) \bigg(
  \frac{9}{2 \sqrt{2}} \Im(F_1 F_2^*) (\cos{\theta_2}-\cos{\theta_1})
  \nonumber
  \\
  & \qquad \qquad \qquad \qquad \qquad \qquad -\frac{9
    \mathsf{u}_2^2}{4} (\cos{\theta_1}+\cos{\theta_2}) \left( \Im(F_3
    F_4^*) \frac{1}{\mathsf{w}} - \sqrt{3} \, \Im(F_L F_M^*)
    \frac{1}{\mathsf{v}}\right) \bigg) \nonumber \\
  & \qquad +(\sin{\theta_1} \sin{\theta_2} \cos{\phi}) \bigg( \Re(F_1
  F_M^*) (\cos{\theta_1}-\cos{\theta_2}) \left(-\frac{9 M_H
      \mathsf{u}_1}{\sqrt{2} \mathsf{v}}\right) \nonumber \\
  & \qquad \qquad \qquad \qquad \qquad \qquad - \frac{9
    \mathsf{u}_2^2}{4} (\cos{\theta_1}+\cos{\theta_2}) \left( \sqrt{3}
    \, \Re(F_3 F_M^*) \frac{1}{\mathsf{v}} - \Re(F_4 F_L^*)
    \frac{1}{\mathsf{w}} \right) \bigg) \nonumber \\
  & + \eta ^2 \Bigg( \frac{9}{4 \mathsf{u}_1^2 \mathsf{v}^2
    \mathsf{w}^2} (\sin{\theta_1} \sin{\theta_2} \cos{\phi}) \bigg(
  \sqrt{2} \mathsf{u}_1^2 \mathsf{v}^2 \mathsf{w}^2 \Re(F_1
  F_2^*)-\mathsf{u}_2^4 \mathsf{v}^2 \mathsf{w} \Re(F_3
  F_4^*)+\sqrt{3}
  \mathsf{u}_2^4 \mathsf{v} \mathsf{w}^2 \Re(F_L F_M^*) \nonumber \\
  & \qquad \qquad \qquad \qquad \qquad \qquad \qquad \qquad + \sqrt{q_1^2 \, q_2^2}
  \left( \mathsf{v}^2 \mathsf{w}^2 \left( \modulus{F_3}^2 -
      \modulus{F_L}^2 \right) - \mathsf{u}_2^4 \left( \modulus{F_4}^2
      \mathsf{v}^2 - 3 \modulus{F_M}^2 \mathsf{w}^2 \right) \right)
  \bigg) \nonumber \\
  & \qquad +\frac{9}{4 \mathsf{u}_1^2 \mathsf{v} \mathsf{w}}
  (\sin{\theta_1} \sin{\theta_2} \sin{\phi}) \bigg( 2 \sqrt{2} M_H
  \mathsf{u}_1^3 \mathsf{w} \Im(F_1 F_M^*)+2 \sqrt{q_1^2 \, q_2^2} \mathsf{v}
  \mathsf{w} \Im(F_3 F_L^*) \nonumber \\
  & \qquad \qquad \qquad +\mathsf{u}_2^4 \left(-\sqrt{3} \mathsf{w}
    \Im(F_3 F_M^*)-\mathsf{v} \Im(F_4 F_L^*)-2 \sqrt{3} \sqrt{q_1^2 \, q_2^2}
    \Im(F_4
    F_M^*)\right) \bigg) \nonumber \\
  & \qquad +\frac{9}{4} \cos{\theta_1} \cos{\theta_2} \bigg(
  -\modulus{F_2}^2+\modulus{F_4}^2 \frac{2 M_H^2
    \mathsf{u}_1^2}{\mathsf{w}^2} \nonumber \\
  & \qquad \qquad \qquad \qquad \qquad \qquad -\modulus{F_M}^2
  \frac{\mathsf{u}_1^2}{\mathsf{v}^2 \mathsf{w}^2 X^2} \left(2 M_H^6
    \mathsf{u}_1^2-M_H^4 \left(3 q_1^2+q_2^2\right) \left(q_1^2+3
      q_2^2\right)+\mathsf{u}_2^8\right) \bigg)
  \Bigg). \label{eq:eta-etasq}
\end{align}

\end{widetext}

\end{document}